\documentclass[11pt]{article}

\usepackage{amssymb}
\usepackage{amsmath}
\usepackage{epsfig}

\newtheorem{thm}{Theorem}
\newtheorem{lem}{Lemma}

\newenvironment{cor}{\medskip\noindent{\bf Corollary:} }{\medskip}
\newenvironment{pf}{\noindent{\bf Proof:} }{\hfill$\Box$\medskip}

\newenvironment{exmp}{\medskip\noindent{\bf Example:} }{\medskip}

 \usepackage{epsfig}

\usepackage{amssymb}
\usepackage{amsmath}

\newcommand{\qed}{}
\renewcommand{\mod}{{{\rm \,mod\,}}} 

\newcommand{\set}[1]{\{#1\}}

\newcommand{\genera}{\stackrel{\star}{\Rightarrow}} 
\newcommand{\Genera}{\stackrel{{\scriptscriptstyle +}}{\Rightarrow}} 

\newcommand{\symb}{\mbox{\rm Symb}}
\newcommand{\var}{\mbox{\rm Var}}
\newcommand{\generaP}{\stackrel{\star}{\Rightarrow}_{G'}} 
\newcommand{\GENERAP}[1]{\stackrel{#1}{\Rightarrow}_{G'}}
\newcommand{\generaG}{\stackrel{\star}{\Rightarrow}_G} 
\newcommand{\GENERAG}[1]{\stackrel{#1}{\Rightarrow}_G}
\newcommand{\RightP}{{\Rightarrow}_{G'}}

\begin{document}
\title{Descriptional Complexity of Bounded Context-Free Languages\footnotemark[1]
\footnotetext[1]{A preliminary version of this work was presented at the 11th Int. Conf. Developments in Language Theory, Turku, Finland, 
July 3-6, 2007.%
}}%
\author{%
Andreas Malcher\footnotemark[2]\and
Giovanni Pighizzini\footnotemark[3]\and
\mbox{}\\
{\normalsize \mbox{\footnotemark[2]\ }\,Institut f\"ur Informatik}\\
{\normalsize Universit\"at Giessen}\\
{\normalsize Arndtstr. 2, 35392 Giessen, Germany}\\
{\normalsize\tt malcher@informatik.uni-giessen.de}\\
\mbox{}\\[-1.1ex]%
{\normalsize \mbox{\footnotemark[3]\ }\,Dipartimento di
    Informatica e Comunicazione}\\
{\normalsize Universit\`{a} degli Studi di Milano}\\
{\normalsize via Comelico 39, 20135 Milano, Italy}\\
{\normalsize\tt pighizzini@dico.unimi.it}%
}%
\date{}%
\maketitle\thispagestyle{empty}%

\begin{abstract}
Finite-turn pushdown automata (PDA) are investigated concerning their descriptional complexity. It is
known that they accept exactly the class of ultralinear context-free languages. Furthermore,
the increase in size when converting arbitrary PDAs accepting ultralinear languages to finite-turn PDAs
cannot be bounded by any recursive function. The latter phenomenon is known as non-recursive trade-off.
In this paper, finite-turn PDAs accepting bounded languages are considered. First, letter-bounded languages are studied.
We prove that in this case the non-recursive trade-off is reduced to a recursive trade-off, more
precisely, to an exponential trade-off. A conversion algorithm is presented and the optimality of the
construction is shown by proving tight lower bounds. Furthermore, the question of reducing the number of
turns of a given finite-turn PDA is studied. Again, a conversion algorithm is provided which shows that
in this case the trade-off is at most polynomial. Finally, the more
general case of word-bounded languages is investigated.
We show how the results obtained for letter-bounded languages can
be extended to word-bounded languages.
\end{abstract}


\section{Introduction}

\label{s:intro}

Finite-turn pushdown automata (PDAs) were introduced in~\cite{ginsburgspanier} by Ginsburg and Spanier.
They are defined by fixing a constant bound on the number of switches between push and pop operations in
accepting computation paths of PDAs. The class of languages defined by these
models is called the class of \emph{ultralinear languages} and is a proper subclass
of the class of context-free languages.
It can be also characterized in terms of \emph{ultralinear} and \emph{non-terminal bounded
grammars}~\cite{ginsburgspanier}.
(In the special case of 1-turn PDAs, i.e., devices making at most
one switch between push and pop operations, we get the class of linear context-free languages).

In~\cite{malcher}, descriptional complexity questions concerning finite-turn PDAs were investigated, by
showing, among other results, the existence of non-recursive trade-offs between PDAs and finite-turn
PDAs. Roughly speaking, this means that for any recursive function $f(n)$ and for arbitrarily large
integers $n$, there exists a PDA of size $n$ accepting an ultralinear language such that any equivalent
finite-turn PDA must have at least $f(n)$ states. Thus, a PDA with arbitrary many turns may represent an
ultralinear language more succinctly than any finite-turn PDA and the savings in size cannot be bounded
by any recursive function.

This phenomenon of non-recursive trade-offs was first observed between con\-text-free grammars and
deterministic finite automata (DFAs) in the fundamental paper by Meyer and Fischer~\cite{meyerfischer}.
Nowadays, many non-recursive trade-offs are known which are summarized, e.g., in \cite{gkklmw} and
\cite{kutrib}. In the context of context-free languages non-recursive trade-offs are known to exist
between PDAs and deterministic PDAs (DPDAs), between unambiguous PDAs (UPDAs) and DPDAs, and between
PDAs and UPDAs. 
Recursive trade-offs are known, e.g., between nondeterministic/alternating finite
automata and DFAs and between DPDAs and DFAs. 

Interestingly, the witness languages used in \cite{meyerfischer} were defined over an alphabet of two
symbols and leave open the unary case which was recently solved in~\cite{pighizzini} by proving an
exponential trade-off. Thus, the non-recursive trade-off in the binary case turns into a recursive
trade-off in the unary case. 
More generally,
a careful investigation of the known cases of non-recursive trade-offs reveals that the used witness
languages are not bounded resp. word-bounded, i.e., they are not included in some subset of $w_1^*w_2^*\ldots w_m^*$ for some fixed words $w_1,w_2,\ldots,w_m$. So, the question arises whether the above non-recursive
trade-offs can be translated to the bounded case or whether the structural limitation on boundedness is
one that will allow only recursive trade-offs.

In this paper, we tackle this question and restrict ourselves initially to the case of letter-bounded languages,
namely, subsets of $a_1^*\ldots a_m^*$, where $a_1,\ldots,a_m$ are pairwise distinct symbols. Our main result shows that
for these languages the trade-off between PDAs (or context-free grammars) and finite-turn PDAs becomes
recursive. More precisely, in Section~\ref{sec:cfg} we first show that each context-free grammar in Chomsky normal form with $h$ variables generating a letter-bounded set can be converted to an equivalent
finite-turn PDA whose size is $2^{O(h)}$. Furthermore, the resulting PDA makes at most $m-1$ turns where
$m$ is the number of letters in the terminal alphabet. In a second step, an exponential trade-off is also shown for arbitrary context-free grammars. 

We prove in Section~\ref{sec:lower} that
this result is tight by showing that the size of the resulting PDA and the number of turns cannot be
reduced. Note that this result is a generalization of the above-mentioned transformation of unary
context-free grammars into finite automata which is presented in~\cite{pighizzini}.
In Section~\ref{sec:turns} the investigation is further deepened by studying how to reduce the number of
turns in a PDA. In particular, given a $k$-turn PDA accepting a subset of $a_1^*a_2^*\ldots a_m^*$,
where $k>m-1$, we show how to build an equivalent $(m-1)$-turn PDA. It turns out that in this case the
trade-off is polynomial. This result is also used to prove the optimality of our simulation of PDAs
accepting letter-bounded languages by finite-turn PDAs. Finally, in Section~\ref{sec:wb}, we consider word-bounded languages. Based on the constructions for letter-bounded languages in the previous sections, we are able to give similar constructions for word-bounded languages. Thus, similar upper and lower bounds can be obtained for the general situation of word-bounded languages. 

We would like to remark that bounded context-free languages have very appealing properties concerning their decidability questions. It is known \cite{ginsburg} that equivalence and inclusion problems are decidable
whereas both problems are undecidable for context-free languages and inclusion is an undecidable problem for deterministic context-free languages. Furthermore, it is decidable whether a given context-free grammar generates a bounded language. In the positive case, the words $w_1,w_2, \ldots, w_m$ can be effectively calculated. For the membership problem we know the Cocke-Younger-Kasami algorithm which solves the problem in cubic time. It is shown in \cite{ibarra} that letter-bounded context-free languages can be accepted by a certain massively parallel computational model. This result implies that the membership problem for letter-bounded context-free languages can be solved in quadratic time and linear space. Since the membership problem for word-bounded context-free languages can be reduced to the membership problem for letter-bounded context-free languages by using suitable inverse homomorphisms, we obtain identical time and space bounds also in the word-bounded case.

\section{Preliminaries and Definitions}

Given a set $S$, $\#S$ denotes its cardinality.
Let $\Sigma^*$ denote the set of all words over the finite alphabet $\Sigma$, with the empty string
denoted by $\epsilon$, and
$\Sigma^+ = \Sigma^* \setminus \{\epsilon\}$. 
Given a string $x\in\Sigma^*$, $|x|$ denotes its length.
For the sake of simplicity, we will consider languages without the empty 
word $\epsilon$. However, our results can be easily extended to 
languages containing $\epsilon$.
Let REG denote the family of regular languages. 
We assume that the reader is familiar with the common notions of formal language theory as presented
in~\cite{hopcroftullman}.

\smallbreak 
A \emph{context-free grammar} (CFG, for short), is a 
4-tuple $G=(V,\Sigma,P,S)$, where $V$ is the set of variables, 
$\Sigma$ is the set of terminals, $V$ and $\Sigma$ are disjoint sets, $S\in V$ is the 
initial symbol and $P\subseteq V\times (V\cup\Sigma)^*$ is the finite 
set of productions. A production $(A,\alpha)\in P$ is 
denoted by $A\rightarrow\alpha$. The 
relations $\Rightarrow$, $\genera$, 
and $\Genera$ are defined in the usual way.
Given $\alpha,\beta\in(V\cup\Sigma)^*$, if
$\theta$ is a derivation of $\beta$ from $\alpha$, then we write 
$\theta:\alpha\genera\beta$.
A useful representation of derivations of context-free grammars can 
be obtained using \emph{parse trees}.

A parse tree (or \emph{tree}, for short) for a context-free 
grammar $G$ is a labeled tree satisfying the following conditions:
\begin{enumerate}
	\item Each internal node is labeled by a variable in $V$.
	\item Each leaf is labeled by either a variable, a terminal, or 
	$\epsilon$. However, if the leaf is labeled $\epsilon$, then it must 
	be the only child of its parent.
	\item If an internal node is labeled with a variable $A$, and its
	children, from left to right, are labeled with $X_1,X_2,\ldots, X_k\in 
	V\cup\Sigma$, then $A\rightarrow X_1X_2\ldots X_k$ is a production 
	of $G$.
\end{enumerate}
If $T$ is a parse tree whose root is 
labeled with a variable $A\in V$ and such that the labels of 
the leaves, from left to right, form a string 
$\alpha\in(V\cup\Sigma)^*$, then we write $T:A\genera\alpha$. Furthermore, we indicate as $\nu(T)$ the
set of variables which appear as labels of some nodes in $T$. 

The language generated by the grammar $G$, i.e., the set 
$\set{x\in\Sigma^*\mid S\genera x}$, is denoted by $L(G)$.

The class of languages generated by CFGs is called the class of
\emph{context-free languages.}
It is well-known that the class of context-free languages properly contains the class of 
regular languages (i.e., the languages accepted by finite automata),
but in the unary case, these two classes collapse \cite{GR62}.

A grammar $G=(V,\Sigma,P,S)$ is said to be in \emph{Chomsky normal form} 
if and only if its productions have the form $A\rightarrow BC$ or
the form $A\rightarrow a$, with $A,B,C\in V$ and $a\in\Sigma$. 
It is well-known that each context-free language not
containing the empty word can be generated by a context-free grammar 
in Chomsky normal form.

A production $A \rightarrow \alpha$ of a context-free grammar $G=(V,\Sigma,S,P)$ is said to be {\em right-linear (left-linear, linear)\/} if $\alpha \in \Sigma^*V$ ($\alpha \in V\Sigma^*$, $\alpha \in \Sigma^*V\Sigma^*$).

A context-free grammar $G=(V,\Sigma,S,P)$ is said to be {\em ultralinear\/} \cite{ginsburgspanier} if
$V$ is a union of pairwise disjoint (possibly empty) subsets $V_0, \ldots, V_n$ of $V$ with the following
property. For each $V_i$ and each $A \in V_i$, each production with left side $A$ is of the form 
$A \rightarrow w$, where either $w \in \Sigma^*V_i\Sigma^*$ or $w \in (\Sigma \cup V_0 \cup \ldots \cup
V_{i-1})^*$. The family of languages generated by ultralinear grammars is called
ultralinear languages and denoted by $\mathrm{ULTRALIN}$.

A context-free grammar $G=(V,\Sigma,S,P)$ is said to be {\em non-terminal bounded\/} \cite{ginsburgspanier} if
there exists an integer $k$ with the following property: If $A \genera w$, $w \in (V \cup \Sigma)^*$, $A \in V$, then $w$ has at most $k$ occurrences of variables. The {\em rank\/} $r_G(w)$ of a word $w \in
(V \cup \Sigma)^*$ is defined to be the largest integer $r$ such that there is a word $u \in (V \cup \Sigma)^*$, with $r$
occurrences of variables, such that $w \genera u$. It is known \cite{ginsburgspanier} that a
context-free grammar $G$ is ultralinear if and only if $G$ is non-terminal bounded. 

For each ultralinear
grammar $G$, the rank of $G$ is defined as the largest integer which is the rank of one of the
variables. Let $L$ be an ultralinear language. The rank of $L$, $r(L)$, is defined as zero, if $L$ is
regular. If $L$ is nonregular, then $r(L)$ is defined as the smallest integer which is
the rank of some ultralinear grammar generating it.

Let $M=(Q,\Sigma,\Gamma,\delta,q_0,Z_0,F)$ be a pushdown automaton \cite{hopcroftullman}. A {\em
configuration\/} of a pushdown automaton is a triple $(q,w,\gamma)$ where $q$ is the current state, $w$
the unread part of the input, and $\gamma$ the current content of the pushdown store. The leftmost
symbol of $\gamma$ is the topmost stack symbol. We write $(q,aw,Z\gamma) \vdash (p,w,\beta\gamma)$, if
$\delta(q,a,Z) \ni (p,\beta)$ for $p,q \in Q$, $a \in \Sigma \cup \{\epsilon\}$, $w \in \Sigma^*$,
$\gamma,\beta \in \Gamma^*$, and $Z \in \Gamma$. The reflexive and transitive closure of $\vdash$ is
denoted by $\vdash^*$. The language accepted by $M$ with accepting states is
$$T(M)=\{w \in \Sigma^* \mid
(q_0,w,Z_0) \vdash^* (q,\epsilon,\gamma) \textrm{ with } q \in F \textrm{ and }\gamma \in \Gamma^*\}.$$

A sequence of configurations of $M$
$(q_1,w_1,\gamma_1)\vdash\ldots\vdash(q_k,w_k,\gamma_k)$ is called {\em one-turn\/} \cite{balkeboehling}
if there exist $1< i \leq j < k$ such that 
\[
|\gamma_1| \leq \cdots \leq |\gamma_{i-1}| < |\gamma_i| \leq |\gamma_{i+1}| \leq \cdots \leq |\gamma_{j}|
> |\gamma_{j+1}| \geq \cdots \geq|\gamma_k|.
\]

A sequence of configurations $c_0 \vdash\ldots\vdash c_m$ is called {\em $k$-turn\/} if there are
integers
$0=i_0,\ldots,i_l=m$ with $l \le k$ such that for $j=0,\ldots,l-1$ the subsequences 
$c_{i_j} \vdash\ldots\vdash c_{i_{j+1}}$ are one-turn, respectively.
$M$ is a $k$-turn pushdown automaton if every word $w \in T(M)$ is accepted by a sequence of
configurations which is $k$-turn. 

By $\mathcal{L}(\textit{k-turn PDA})$ we denote the family of languages accepted by $k$-turn PDAs.
The union of all $k$-turn PDAs with fixed $k \ge 1$ is the set of finite-turn PDAs. The family of
languages accepted is defined as $\mathcal{L}(\textit{finite-turn PDA})=\bigcup_{k \ge 1}
\mathcal{L}(\textit{k-turn PDA})$. 

Thus, $k$-turn PDAs are allowed to make new turns not depending on the stack height.
The following characterization of ultralinear languages by finite-turn PDAs may be found in
\cite{balkeboehling} and \cite{ginsburgspanier},
respectively.

\begin{thm} A language $L$ belongs to $\mathrm{ULTRALIN}$ if and only if there is a $k \in \mathbb{N}$ such that $L$ is accepted by a $k$-turn PDA.
\end{thm}

We want to consider in this paper PDAs in a certain normal form. Thus, we make,
without loss of generality, the following assumptions about PDAs (cf. \cite{pighizzini}).

\begin{enumerate}
\item At the start of the computation the pushdown store contains only the start symbol $Z_0$; this
symbol is never pushed or popped on the stack;
\item the input is accepted if and only if the automaton reaches a final state, the pushdown store only
contains $Z_0$ and all the input has been scanned;
\item if the automaton moves the input head, then no operations are performed on the stack;
\item every push adds exactly one symbol on the stack.
\end{enumerate}

The transition function $\delta$ of a PDA $M$ then can be written as
\[
\delta : Q \times (\Sigma \cup \{\varepsilon\}) \times \Gamma \rightarrow 2^{Q \times
(\{-,pop\}\cup\{push(A) \mid A \in \Gamma\})}.
\]
In particular, for $q,p \in Q, A,B \in \Gamma, \sigma \in \Sigma, (p,-) \in \delta(q,\sigma,A)$ means
that the PDA $M$, in the state $q$, with $A$ at the top of the stack, by consuming the input $\sigma$,
can reach the state $p$ without changing the stack contents. $(p,pop) \in \delta(q,\varepsilon,A)$
($(p,push(B)) \in \delta(q,\varepsilon,A), (p,-) \in \delta(q,\varepsilon,A)$, respectively) means that
$M$, in the state $q$, with $A$ at the top of the stack, without reading any input symbol, can reach the
state $p$ by popping off the stack the symbol $A$ on the top (by pushing the symbol $B$ on the top of
the stack, without changing the stack, respectively).

A {\em descriptional system\/} $D$ is a recursive set of
finite descriptors (e.g. automata or grammars) relating each $A \in D$ to a
language $T(A)$. It is additionally required that each descriptor $A \in D$ can be effectively converted
to a Turing machine $M_A$ such that $T(M_A)=T(A)$. The language family being described by $D$ is
$T(D)=\{T(A) \mid A \in D \}$. For every language $L$ we define $D(L)=\{A \in D \mid T(A)=L\}$. A
{\em complexity measure\/} for $D$ is a total, recursive, finite-to-one function $| \cdot |: D
\rightarrow \mathbb{N}$ such that the descriptors in $D$ are recursively enumerable in order of
increasing complexity. Comparing two descriptional systems $D_1$ and $D_2$,
we assume that $T(D_1) \cap T(D_2)$ is not finite.

We say that a function $f : \mathbb{N} \rightarrow \mathbb{N}$, $f(n) \ge n$ is an {\em upper bound\/}
for the blow-up in complexity when changing from one
descriptional system $D_1$ to another system $D_2$, if every description $A \in D_1$ of size $n$
has an equivalent description $A' \in D_2$ of size at most $f(n)$.

We say that a function $g : \mathbb{N} \rightarrow \mathbb{N}$, $g(n) \ge n$
is a {\em lower bound\/} for the trade-off between two descriptional systems $D_1$ and $D_2$, if there
is an infinite sequence $N \subseteq \mathbb{N}$ and an infinite sequence $(L_n)_{n \in N}$ of pairwise
distinct languages $L_n$ such that for all $n \in N$ there is a description $A \in D_1$ for $L_n$ of
size $n$ and every description $A' \in D_2$ for $L_n$ is at least of size $g(n)$.

According to the discussion in \cite{harrison} the size of a PDA should be defined depending on the
number of states, the number of stack symbols, the number of input symbols, and the maximum number
of stack symbols appearing in the right hand side of transition rules. In this paper, we consider PDAs in
the above defined normal form over a fixed alphabet $\Sigma$. Thus, $size(M)$ of a PDA $M$ in normal
form is defined as the product of the number of states and the number of stack symbols.
It can be observed that this measure fulfills the above defined conditions on descriptional measures.
The size of a finite automaton is defined to be the number of states.

As a measure for the size of a context-free grammar $G=(V,\Sigma,P,S)$ we consider the number of symbols of $G$, defined as $\symb(G)=\sum_{(A\rightarrow\alpha)\in P}(2+|\alpha|)$ (cf. \cite{kelemenova}).
Furthermore, in the paper it will be useful also to consider
the number of variables of $G$, defined as $\var(G)=\#V$ (note
that this function in general is not a measure for the size).
Some general information on descriptional complexity may be found in \cite{gkklmw}.

\section{From Grammars to Finite-Turn PDAs}
\label{sec:cfg}

In this section, we study the transformation of CFGs into finite-turn PDAs.
Our main result shows that given a grammar $G$ of size $h$,
we can build an equivalent finite-turn PDA $M$ of size $2^{O(h)}$.
Furthermore, if the terminal alphabet of $G$ contains $m$ letters, then
$M$ is an $(m-1)$-turn PDA. The tightness of the bounds will be shown 
in Section~\ref{sec:lower}.
For the sake of simplicity, we start by considering CFGs in Chomsky
normal form with the measure $\var$. At the end of the
section, we will discuss the generalization to arbitrary context-free
grammars, taking into consideration the size defined by the measure
$\symb$.

In the following we consider an alphabet 
$\Sigma=\set{a_1,\ldots,a_m}$ and a CFG $G=(V,\Sigma,P,S)$ 
in Chomsky normal form with $h$ variables, generating a subset of 
$a_1^*\ldots a_m^*$. Without loss of generality, we can suppose that each variable
of $G$ is \emph{useful}, i.e., for each $A\in V$, there exist terminal strings
$u,v,w$, such that $S\genera uAw\genera uvw$.

\begin{lem}\label{lemma:A}
  For each variable $A\in V$ there exists an index $l$, $1\leq l\leq m$
  ($r$, $1\leq r\leq m$, resp.) such that if $A\Genera uAv$, with 
  $u,v\in\Sigma^*$, then $u\in a_l^*$  ($v\in a_r^*$, resp.).
  Furthermore, if there exists at least one derivation $A\Genera uAv$
  with $u\neq\epsilon$, ($v\neq\epsilon$, resp.) then such an $l$ ($r$, resp.)
  is unique.
\end{lem}

\begin{pf}
  It is easy to see that if $A\Genera uAv$ and $u$ contains at least two
  letters $a_{l'},a_{l''}$, with $l'\neq l''$, then, because
  $A\Genera uuAvv$, the language generated by $G$ should contain a string not
  belonging to $a_1^*\ldots a_m^*$.
  Hence, $u\in a_l^*$, for some $1\leq l\leq m$.
  
  Now suppose that $u\neq\epsilon$ and $A\Genera u'Av'$.
  Because $A\Genera uu'Av'v$, using the previous argument it is easy to 
  conclude that $u'\in a_l^*$.
  
  A similar argument can be given for the right part.
\qed
\end{pf}

For each variable $A$ such that $A\Genera uAv$, with $u,v\in\Sigma^+$, we
denote by $border(A)$ the unique pair $(l,r)$ of indices, given in Lemma~\ref{lemma:A}. On the other
hand, if for any derivation $A\Genera uAv$,
the string $v$ is empty, then we define $border(A)$ as the pair $(l,l)$, and
if for any derivation $A\Genera uAv$ the string $u$ is empty, then we define $border(A)$ as the pair
$(r,r)$.
If there are no derivations of the form $A\Genera uAv$, then we leave $border(A)$ undefined. 

Formally,
\[
border(A)=\left\{
\begin{array}{ll}
	(l,r)&\mbox{if $A\Genera uAv$ for some 
                       $u\in a_{l}^+, v\in a_{r}^+$}\\
	(l,l)&\mbox{if $u\in a_{l}^+$ and $v=\epsilon$ 
                       for any $A\Genera uAv$}\\
	(r,r)&\mbox{if $u=\epsilon$ and $v\in a_{r}^+$ 
                       for any $A\Genera uAv$}\\
    \mbox{undefined }&\mbox{otherwise}.
\end{array}\right.
\]
For the sake of brevity, $border(A)$ will be denoted also as $(l_A,r_A)$,
if defined.

\medskip

We now consider the relation $\leq$ on the set of possible borders
defined as $(l,r)\leq (l',r')$ if and only if $l\leq l'$ and if $l=l'$ then $r\geq r'$, for all
$(l,r),(l',r')\in\set{1,\ldots,m}^2$, with $l\leq r$ and $l'\leq r'$.
It is not difficult to verify that $\leq$ is a total order on the set of pairs of indices $l,r$ from
$\set{1,\ldots,m}$, such that $l\leq r$.

Actually, we are interested in computing borders of variables belonging to the same derivation tree. In
this case, either a variable is a descendant of the other in the tree, and then the interval defined by
its border is inside the interval defined by the border of the other variable, or one variable is to the
right of the other one, and then the corresponding interval is to the right of the other one. More
formally, we can prove the following:

\begin{lem}
\label{lemma:border}
  Let $T$ be a derivation tree and $A,B\in\nu(T)$ be two variables. If
  $border(A)\leq border(B)$, then either: 
  
  (a) $l_A\leq l_B\leq r_B\leq r_A$, or
  
  (b) $l_A< l_B$, $r_A<r_B$, and $r_A\leq l_B$.

\end{lem}

\begin{pf}
  The case $border(A)=border(B)$ is trivial. Thus, for the rest of the
  proof we suppose that $border(A)\neq border(B)$.
  
  If $A$ and $B$ lie in $T$ on the same path from the root, then $A$ must
  be closer to the root than $B$ (otherwise $border(B)\leq border(A)$). 
  It is immediate
  to conclude that in this case $l_A\leq l_B\leq r_B\leq r_A$.

  Otherwise, $A$ should appear in a path to the left of the path containing
  $B$. This easily implies that $l_A\leq r_A\leq l_B\leq r_B$.
  We consider the two subcases $r_A=r_B$ and $r_A<r_B$.
  
  If $r_A=r_B$ then we get $l_A<l_B=r_B=r_A$, so (a) holds.
  On the other hand, if $r_A<r_B$ then it is not possible that $l_A=l_B$
  because this should imply, with $border(A)\leq border(B)$, the contradiction 
  $r_B\leq r_A$. Hence, $l_A<l_B$ that, with $r_A<r_B$ and $r_A\leq l_B$,
  gives (b).
%
\qed
\end{pf}

A \emph{partial derivation tree} (or \emph{partial tree}, for short) $U: A\genera vAx$
is a parse tree whose root is labeled with a variable $A$ and all the
leaves, with the exception of one whose label is the same variable $A$, are labeled with terminal
symbols.

Given a partial tree $U: A\genera vAx$, any derivation tree $T:S\genera z$
with $A\in\nu(T)$ can be ``pumped'' using $U$, by replacing a node
labeled $A$ in $T$ with the subtree $U$.
In this way, a new tree $T':S\genera z'$ is obtained, where $z'=uvwxy$,
such that $z=uwy$, $S\genera uAy$, and $A\genera w$. Moreover,
$\nu(T')=\nu(T)\cup\nu(U)$.

On the other hand, any derivation tree producing a sufficiently long terminal string can be obtained by
pumping a derivation tree of a shorter string with a partial tree.
This fact, which is essentially the
pumping lemma of context--free languages (see, e.g., \cite{hopcroftullman}), 
is recalled in the following:

\begin{lem}
\label{lemma:pump}
  Let $T:S\genera z$ be a derivation tree of a string $z\in\Sigma^*$.
  If $|z|>2^{h-1}$ then we can write $z=uvwxy$, such that 
  $0<|vx|<2^h$, there is a tree $T':S\genera uwy$, a variable 
  $A\in\nu(T')$ and a tree $T'':A\Genera vAx$ such that
  $\nu(T)=\nu(T')\cup\nu(T'')$ and $A\Genera w$, where $h$ is the number 
  of the variables of the grammar $G$.
\end{lem}
\begin{pf}
  First of all, we recall that given a parse tree of a string $\alpha$
  according to a context-free grammar $G$ in Chomsky normal form, if 
  the longest path from the root to a leaf in the tree has length $k$ 
  (measured by the number of edges), then $|\alpha|\leq 2^{k-1}$
  (see, e.g.,\cite{hopcroftullman}).
  Using this property, we get that if $|z|>2^{h-1}$ then the tree
  $T:S\genera z$ must contain a path of $h+1$ edges from a node $n$
  to a leaf. Hence, this path must contain two nodes labeled with
  the same variable $A$. This defines the decomposition of $T$ in $T'$
  and $T''$. Again by the above property, the terminal
  string $vwx$ generated by the subtree rooted at $n$ has length bounded
  by $2^h$. 
  Furthermore, $|w|\geq 1$, because the grammar is in Chomsky normal form.
  Hence $|vx|<2^h$.
\qed
\end{pf}

By applying the pumping lemma several times, we can prove that any derivation tree can be obtained by starting from a derivation tree of a ``short'' string
(namely, a string of length at most $2^{h-1}$) and iteratively
pumping it with ``small'' partial trees. Furthermore, a sequence of
partial trees can be considered such that the sequence of 
borders of their roots is not decreasing. This fact will be crucial
to get the main simulation presented in this section. More precisely,
we are able to prove the following result:

\begin{lem} \label{lemma:der}	
  Given a derivation tree $T:S\genera z$ of a string $z\in\Sigma^*$,
  with $|z|>2^{h-1}$, for some integer $k>0$ there are:
  \begin{itemize}
  \item
  $k+1$ derivation trees $T_0,T_1,\ldots,T_k$,
  where $T_i:S\genera z_i$, $i=0,\ldots,k$,  $0<|z_0|\leq 2^{h-1}$,
  $T_k=T$, $z_k=z$;
  \item
  $k$ partial trees $U_1,\ldots U_k$, where, for $i=1,\ldots,k$,
  $U_i:A_i\Genera v_iA_ix_i$, $0<|v_ix_i|<2^h$, and $T_i$ is obtained
  by pumping $T_{i-1}$ with $U_i$.
  \end{itemize}
  Furthermore,
  $border(A_1)\leq border(A_2)\leq\ldots\leq border(A_k).$
\end{lem}
\begin{pf}
  We can build the sequence from the end starting from $T$ and
  decomposing it in a tree $T'$ and a partial tree $U$, 
  according to Lemma~\ref{lemma:pump}. 
  This process can be iterated until a derivation  tree
  $T_0$ producing a string of length bounded by $2^{h-1}$ is obtained.

  In order to get a sequence of partial trees such that the sequence  of the
  borders of the variables labeling their roots is not decreasing,  at each step
  a partial tree is selected, among all possible candidates, in such a
  way that the border of its root is maximum. In other words,
  for $i=1,\ldots,k$, the variable $A_i$ labeling the root of the tree
  $U_i:A_i\Genera v_iA_ix_i$ satisfies:
  \begin{eqnarray*}
  \lefteqn{} border(A_i)=\max\{border(A)\mid\hbox{\rm there exists a
          partial tree}\\
      U:A\Genera vAx \hbox{ \rm  of } T_i \hbox{ \rm with } 0<|vx| <2^h\}.
  \end{eqnarray*}
  We prove that with this choice $border(A_i)\leq border(A_{i+1})$,
  for $i=1,\ldots, k-1$. This is obvious if
  the partial tree $U_i:A_i\Genera v_iA_ix_i$ of $T_i$
  is also a partial tree of $T_{i+1}$, namely, 
  it belongs to the set of candidates
  when $U_{i+1}$ is  chosen to reduce $T_{i+1}$.
  If this is not the case, then $T_{i+1}$ should contain
  a partial tree $U':A_i\Genera v'A_ix'$, with $|v'x'|\geq  2^h$, such that,
  after removing $U_{i+1}$, $U'$ is reduced in $T_i$ to $U_i$. It  can be
  observed that in $T_{i+1}$ such a reduction is possible only
  if the root $A_{i+1}$ of
  $U_{i+1}$ is a descendant of the root $A_i$ of $U'$.
  Hence, in $T_{i+1}$ the terminal string generated by the subtree  whose
  root, labeled $A_{i+1}$, coincides with the root of $U_{i+1}$,  must be
  a factor of the terminal string generated by the subtree whose
  root, labeled $A_{i}$, coincides with the root of $U'$.
  This implies that
  $l_{A_i}\leq l_{A_{i+1}}\leq r_{A_{i+1}}\leq r_{A_i}$.
  Hence, $border(A_i)\leq border(A_{i+1})$.
\qed
\end{pf}

\begin{exmp}	
  The language $L=\set{a_1^{n+k}a_2^{k+p}a_3^{p+n}\mid n,k,p > 0}$ can 
  be generated by a grammar in Chomsky normal form with the following
  productions:
  \[
  \begin{array}{lllllllll}
  S\rightarrow A_1E&\hspace*{3mm}&S'\rightarrow AB&\hspace*{3mm}&A\rightarrow A_1F&\hspace*{3mm}&B\rightarrow A_2G&\hspace*{3mm}&A_1\rightarrow a_1\\
  E\rightarrow SA_3&&A\rightarrow A_1A_2&&F\rightarrow AA_2&&G\rightarrow BA_3&&A_2\rightarrow a_2\\
  E\rightarrow S'A_3&&B\rightarrow A_2A_3&&&&&&A_3\rightarrow a_3
  \end{array}
  \]
Note that $S\Genera a_1Sa_3$, $A\Genera a_1Aa_2$, and $B\Genera a_2Ba_3$.
It is easy to get a tree $T_0:S\Genera a_1^2a_2^2a_3^2$ and
three partial trees $U':S\Genera a_1Sa_3$, $U'':A\Genera a_1Aa_2$, and
$U''':B\Genera a_2Ba_3$. 
Given integers $n,k,p>0$, a derivation tree for the string 
$a_1^{n+k}a_2^{k+p}a_3^{p+n}$ can be obtained considering $T_0$,
and pumping it $n-1$ times with the tree $U'$, $k-1$ times with the
tree $U''$, and $p-1$ with the tree $U'''$. Note that
$border(S)=(1,3)\leq border(A)=(1,2)\leq border(B)=(2,3)$.
\end{exmp}

Lemma~\ref{lemma:der} suggests a nondeterministic procedure which can
be used to generate all the strings $a_1^{k_1}\ldots a_m^{k_m}$
belonging to the language $L(G)$: 
at the beginning a derivation tree $T:S\genera a_1^{n_1}\ldots a_m^{n_m}$
of a ``short'' string is selected.
Then the procedure enters a loop which is repeated a nondeterministically
chosen number of times.
At each iteration, the tree $T$ so far considered is pumped with a
nondeterministically chosen partial tree $U:A\Genera vAx$
(note that, by Lemma~\ref{lemma:A}, $v\in a_{l_A}^*$, $x\in a_{r_A}^*$)
such that $A$ is a variable occurring in $T$. Note that, to implement
this strategy, the procedure does not need to remember the whole tree
$T$ but only the set of variables occurring in it.

The procedure is the following:

{
\footnotesize
\begin{tabbing}
nondeterministically select a tree 
$T:S\genera a_1^{n_1}a_2^{n_2}\ldots a_m^{n_m}$,\\
~~~~with $n_1+n_2+\ldots+n_m\leq 2^{h-1}$\\
$k_1\leftarrow n_1$, $k_2\leftarrow n_2$, $\ldots$, $k_m\leftarrow n_m$\\
{\sl enabled }$\leftarrow\nu(T)$\\
{\sl iterate }$\leftarrow$ nondeterministically choose {\sl true }or {\sl 
false}\\
{\bf whi}\={\bf le} {\sl iterate }{\bf do}\\
\>nondeterministically select a tree $U:A\Genera vAx$, with $0<|vx|<2^h$,\\
\>~~~~and $A\in${\sl enabled}~~~~~~~~// $border(A)=(l_A,r_A)$\\
\>$k_{l_A} \leftarrow k_{l_A} + |v|$\\
\>$k_{r_A} \leftarrow k_{r_A} + |x|$\\
\>{\sl enabled }$\leftarrow${\sl enabled }$\cup\ \nu(U)$\\
\>{\sl iterate }$\leftarrow$ nondeterministically choose {\sl true }or {\sl 
false}\\
{\bf endwhile}\\
output $a_1^{k_1}a_2^{k_2}\ldots a_m^{k_m}$\\
\end{tabbing}
}

Now, we will convert the above procedure into an $(m-1)$-turn PDA recognizing
the language generated by the grammar $G$. 
For the sake of simplicity, let us start by describing the case $m=2$
with $v\in a_1^*$ and $x\in a_2^*$, for each partial tree $U:A\Genera vAx$.
The PDA uses two bounded counters $n_1, n_2$ in order to remember
the string $a_1^{n_1}a_2^{n_2}$ generated by the initial ``small'' tree.
In a preliminary phase, the PDA consumes $n_1$ occurrences of $a_1$
from the input tape, in order to verify that $a_1^{n_1}$ is a prefix of
the input (otherwise it stops and rejects).
Subsequently, the automaton starts the simulation of the loop
above described where, at each iteration, a partial tree
$U:A\Genera vAx$, with $A\in{\sl enabled }$ is used to pump the generated
string.
To this aim, the automaton reads $v\in a_1^*$ from the input tape
and pushes $x\in a_2^*$ on the stack (if $v$ is not a prefix of the
remaining part of the input tape, then the automaton stops and rejects).
At the end of the loop, the pushdown store will contain a string
$a_2^p$, for some $p\geq 0$. Finally, the automaton accepts
if and only if the remaining part of the input is $a_2^{n_2+p}$.
We can observe that the automaton so described simulates the
derivation of a string and verifies its matching with the
input string. For the occurrences of the letter $a_1$, the matching
is verified immediately, by comparing the generated factors with the
input string; for the occurrences of the letter $a_2$, the verification
of the matching is postponed: the generated factors are kept on the 
stack and compared with the input in the final phase.

This strategy can be extended to the general case by pumping,
according to Lemma~\ref{lemma:der}, with partial trees such
that the sequence of the borders of their roots is not decreasing.
More precisely, the PDA implements the following
nondeterministic procedure, whose correctness is proved in Lemma~\ref{lemma:pda} and Theorem~\ref{theorem:3-5}.
\pagebreak[4]

{
\footnotesize
\begin{tabbing}
nondeterministically select a tree 
$T:S\genera a_1^{n_1}a_2^{n_2}\ldots a_m^{n_m}$,\\
~~~~with $n_1+n_2+\ldots+n_m\leq 2^{h-1}$\\
read $a_1^{n_1}$ from the input\\
{\sl enabled }$\leftarrow\nu(T)$\\
$(l,r) \leftarrow (1,m)$  // the ``work context''\\
{\sl iterate }$\leftarrow$ nondeterministically choose {\sl true }or {\sl 
false}\\
{\bf whi}\={\bf le} {\sl iterate }{\bf do}\\
\>nondeterministically select a tree $U:A\Genera vAx$, with $0<|vx|<2^h$,\\
\>~~~~$A\in${\sl enabled}, and $(l,r)\leq border(A)=(l_A,r_A)$\\
\>{\bf if }\= $r<r_A$ {\bf then }
      //new context to the right of the previous one\\
\>\>{\bf for} \=$j\leftarrow l+1$ {\bf to }$r-1$ {\bf do}\\
\>\>\>      consumeInputAndCounter($j$)\\
\>\>{\bf endfor}\\
\>\>{\bf for} \=$j\leftarrow r$ {\bf to }$l_A$ {\bf do}\\ 
\>\>\>      consumeInputAndCounter($j$)\\
\>\>\>      consumeInputAndStack($j$)\\
\>\>{\bf endfor}\\
\>{\bf else} //$r_A\leq r$: new context inside the previous one\\
\>\>{\bf for} \=$j\leftarrow l+1$ {\bf to }$l_A$ {\bf do}\\
\>\>\>      consumeInputAndCounter($j$)\\
\>\>{\bf endfor}\\
\>{\bf endif}\\
\>$(l,r)\leftarrow(l_A,r_A)$\\
\>read $v$ from the input\\
\>{\bf if }$r \neq l$ \={\bf then} push $x$ on the stack\\
\>\>{\bf else} read $x$ from the input\\
\>{\bf endif}\\
\>{\sl enabled }$\leftarrow${\sl enabled }$\cup\ \nu(U)$\\
\>{\sl iterate }$\leftarrow$ nondeterministically choose {\sl true }or {\sl 
false}\\
{\bf endwhile}\\
{\bf for} \=$j\leftarrow l+1$ {\bf to } $r-1$ {\bf do}\\
\>      consumeInputAndCounter($j$)\\
{\bf endfor}\\
{\bf for} \=$j\leftarrow r$ {\bf to } $m$ {\bf do}\\
\>     consumeInputAndCounter($j$)\\
\>     consumeInputAndStack($j$)\\
{\bf endfor}\\
{\bf if} the end of the input has been reached \={\bf then} accept\\
\>{\bf else} reject\\
{\bf endif}
\end{tabbing}
}
\noindent
In the previous procedure and in the following macros,
the instruction ``read $x$ from the input tape,'' for 
$x\in\Sigma^*$, actually means that the automaton verifies whether or not 
$x$ is a prefix of the next part of the input. If the outcome of this test
is positive, then the input head is moved immediately to the right of $x$, namely, $x$ is ``consumed,''
otherwise the machine stops and rejects.

\smallskip
\noindent
The macros are defined as follows:

{
\footnotesize
\begin{tabbing}
ConsumeInputAndCounter($j$):\\
{\bf whi}\={\bf le} $n_j\geq 0$ {\bf do}\\
\>read $a_j$ from the input tape\\
\>$n_j\leftarrow n_j-1$\\
{\bf endwhile}\\
\end{tabbing}
}
\pagebreak[4]
{
\footnotesize
\begin{tabbing}
ConsumeInputAndStack($j$):\\
{\bf whi}\={\bf le} the symbol at the top of the stack is $a_j$ {\bf do}\\
\>read $a_j$ from the input tape\\
\>pop\\
{\bf endwhile}\\
\end{tabbing}
}

In order to prove that the pushdown automaton described in the previous
procedure accepts the language $L(G)$ generated by the given grammar $G$,
it is useful to state the following lemma:

\begin{lem}\label{lemma:pda}
  Consider one execution of the previous procedure.
  Let $T_0:S\genera a_1^{n_1}\ldots a_m^{n_m}$ be the tree selected at the
  beginning of such an execution.
  At every evaluation of the condition of the while loop, there exists
  a tree $T:S\genera a_1^{k_1}\ldots a_m^{k_m}$, for some
  $k_1,\ldots,k_m\geq 0$, such that
  \begin{itemize}
  \item the scanned input prefix is $z=a_1^{k_1}\ldots a_l^{k_l}$;
  \item the pushdown store contains the string 
    $\gamma=a_r^{p_r}\ldots a_m^{p_m}$, where, for $j=r,\ldots,m$,
    $p_j\geq 0$ and $k_j=p_j+n_j$;
  \item for $l<j<r$,  $k_j=n_j$;
  \item {\sl enabled }$=\nu(T)$.
  \end{itemize}
\end{lem}

\begin{pf}
  It is easy to see that at the first evaluation of the condition
  it holds that $l=1$, $r=m$, $k_1=n_1,\ldots,k_m=n_m$, $p_m=0$, the
  scanned input prefix is $a_1^{k_1}$ and the pushdown store is empty,
  namely, it contains $a_m^{p_m}$. 
   
  We now suppose the statement to be true before the execution of one iteration and
  we show that it still holds true at the end of the iteration.
  Let $U:A\Genera vAx$ be the partial tree selected in the while loop.
  Because $A\in\nu(T)$, the derivation tree 
  $T:S\genera a_1^{k_1}\ldots a_m^{k_m}$ can be pumped with the partial tree
  $U$, obtaining a new tree $T':S\genera a_1^{k'_1}\ldots a_m^{k'_m}$,
  with $\nu(T')=\nu(T)\cup\nu(U)$, where (in the case $l_A\neq r_A$)
  $k'_{l_A}=k_{l_A}+|v|$, $k'_{r_A}=k_{r_A}+|x|$ and $k'_j=k_j$
  for each $j\neq l_A$ and $j\neq r_A$ (in the special case $l_A=r_A$
  we have that $k'_{l_A}=k'_{r_A}=k_{l_A}+|vx|$).
  
  We now consider two subcases, corresponding to the selection in the
  while loop.

  \medskip\noindent
  \emph{Case $r<r_A$.}\\
  By Lemma~\ref{lemma:border}, this implies that $l<l_A$ and $r\leq l_A$.
  
  First, we prove that for each $j$, with $l_A<j<r_A$, the stack cannot
  contain the symbol $a_j$, i.e., $p_j=0$. Suppose, by contradiction, that
  the string $\gamma$ contains at least one occurrence of $a_j$. This
  symbol must have been pushed on the stack in a previous iteration, with 
  ``work context'' $(\tilde{l},\tilde{r})$, for some $\tilde{l}\leq\tilde{r}=j$.
  Since the procedure never removes variables from the tree, the variable
  used to pump the tree in such a previous iteration is also in the tree $T'$.
  Moreover, the procedure chooses contexts in a nondecreasing order,
  so  $(\tilde{l},\tilde{r})\leq(l_A,r_A)$.
  By Lemma~\ref{lemma:border} it turns out that either 
  $\tilde{l}\leq l_A\leq r_A\leq\tilde{r}=j$, or $\tilde{l}<l_A$,
  $\tilde{r}<r_A$, and $j=\tilde{r}\leq l_A$. It is easy to observe that in
  both the cases we get a contradiction.
  Hence, for any $j$ with $l_A<j<r_A$, the stack does not contain the
  symbol $a_j$, i.e., $p_{l_A+1}=p_{l_A+2}=\ldots=p_{r_A-1}=0$.
  This implies that $\gamma=\gamma''\gamma'$ with 
  $\gamma''=a_r^{p_r}\ldots a_{l_A}^{p_{l_A}}$ and 
  $\gamma'=a_{r_A}^{p_{r_A}}\ldots a_m^{p_m}$.

  Now, we observe the operations on the input and on the stack that are
  performed during the execution of the body of the loop:
  \begin{itemize}
  \item the input factor 
  $a_{l+1}^{n_{l+1}}\ldots a_{r-1}^{n_{r-1}}a_r^{p_r+n_r}\ldots
  a_{l_A}^{p_{l_A}+n_{l_A}}=a_{l+1}^{k_{l+1}}\ldots a_{l_A}^{k_{l_A}}$
  is consumed;
  \item the string $\gamma''$ is popped off the stack;
  \item the input factor $v\in a_{l_A}^*$ is consumed;
  \item if $l_A<r_A$ then the string $x\in a_{r_A}^*$ is saved on the
  stack (to be consumed later), otherwise it is consumed immediately.
  \end{itemize}
  By summarizing, in the case $l_A<r_A$, at the end of the iteration
  the scanned input prefix is 
  $a_1^{k_1}\ldots a_{l_A}^{k_{l_A}}v=a_1^{k'_1}\ldots a_{l_A}^{k'_{l_A}}$, the
  pushdown store contains the string $xa_{r_A}^{p_{r_A}}\ldots a_m^{p_m}$,
  for each $j$, $l_A<j<r_A$, $k'_j=k_j=n_j$, and $enabled=\nu(T')$.
  With small changes we can deal with the case $l_A=r_A$.
  
  \medskip\noindent
  \emph{Case $r_A\leq r$.}\\
  By Lemma~\ref{lemma:border}, this implies that $l\leq l_A\leq r_A\leq r$.
  
  If $l_A<r_A$  then the consumed input prefix is 
  $a_1^{k_1}\ldots a_{l_A}^{k_{l_A}}v$ and the pushdown store contains
  the string $x\gamma = xa_{r_A}^{p_{r_A}}\ldots a_m^{p_m}$, where
  $p_{r_A}=p_{r_A+1}=\ldots=p_{r-1}=0$. At this point it
  is not difficult to verify that the statement of the Lemma is true.
  The subcase  $l_A=r_A$ can be managed with easy changes.
\qed
\end{pf}

As a consequence:

\begin{thm}\label{theorem:3-5}
  The pushdown automaton $M$ described by the previous procedure is an 
  $(m-1)$--turn PDA accepting the language $L(G)$.
\end{thm}

\begin{pf}
  First, we show that the number of turns of the PDA $M$ defined in the above
  procedure is at most $m-1$. To this aim we count how many times the
  automaton can switch from push operations to pop operations.
  
  At each iteration of the while loop, the automaton can perform 
  push operations.
  Pop operations are possible only by calling the macro consumeInputAndStack. 
  This happens first in the while loop, when the condition
  $r<r_A$ holds true, i.e., when the new context $(l_A,r_A)$ is to
  the right of the previous context $(l,r)$, and secondly after the end of the loop.

  Let $(l_1,r_1), (l_2,r_2), \ldots (l_k,r_k)$ be the sequence of the contexts
  which in the computation make the above-mentioned condition hold true. Hence,
  $1<l_1<\ldots<l_k\leq m$, that implies $k\leq m-1$.
  If $k<m-1$, then the PDA $M$ makes at most
  $k\leq m-2$ turns in the simulation of the while loop and one more 
  turn after the loop.
  So the total number of turns is bounded by $m-1$.
  
  Now, suppose that $k=m-1$. This implies that $l_k=m=r_k$. Before reaching
  the context $(l_k,r_k)$, at most $m-2$ turns can be performed. 
  When the automaton
  switches to the new context $(l_k,r_k)=(m,m)$, it can make pop operations, by
  calling the macro consumeInputAndStack($m$). This requires one more turn.
  After that, the automaton can execute further iterations, 
  using the same context
  $(m,m)$. By reading the procedure carefully, 
  we can observe that it never executes 
  further push operations. Finally, at the exit of the loop, 
  further pop operations
  can be executed (consumeInputAndStack). Hence, the total number of turns is
  bounded by $m-1$.
  
  \medskip
  To prove that the language $L(G)$ and the language accepted by the
  automaton defined in the above procedure coincide it is enough to
  observe that given a string $z\in L(G)$, the procedure is able to guess
  the tree $T_0$ and the partial trees $U_1,\ldots, U_k$ of 
  Lemma~\ref{lemma:der}, recognizing in this way $z$.
  Conversely, using Lemma~\ref{lemma:pda}, it is easy to show that each 
  string accepted by the procedure should belong to $L(G)$.
\qed
\end{pf}

\begin{cor}\label{cor:cfg}
  Given an alphabet $\Sigma=\set{a_1,\ldots,a_m}$, for any context--free
  grammar $G$ in Chomsky normal form with $h$ variables generating 
  a letter-bounded language $L\subseteq a_1^*\ldots a_m^*$, there exists an equivalent
  $(m-1)$-turn PDA $M$ with $2^{O(h)}$ states and $O(1)$ stack symbols.
\end{cor}

\begin{pf}
  The most expensive information that the automaton defined in the 
  previous procedure has to remember in
  its state are the $m-1$ counters bounded by $2^{h-1}$, 
  and the set $enabled$, which  is a subset of $V$. For the pushdown store
  an alphabet with $m+1$ symbols can be used. With a small modification,
  the pushdown store can be implemented using only two symbols (one symbol
  to keep a counter $p_j$ and another one to separate two consecutive
  counters), and increasing the number of states by a factor $m$, to remember
  what input symbol $a_j$ the stack symbol $A$ is representing.
\qed
\end{pf}

Using standard techniques, a PDA of size $n$ can be converted to an equivalent CFG in Chomsky normal form
with $O(n^2)$ variables. 
Hence, we easily get:

\begin{cor}\label{cor:pda}
  Each PDA of size $n$ accepting a subset of $a_1^*\ldots a_m^*$ can be
  simulated by an equivalent $(m-1)$-turn PDA of size $2^{O(n^2)}$.
\end{cor}

We now consider the situation when the given CFG is not necessarily in Chomsky normal form.

\begin{lem}\label{lemma:norm}
  Given a context-free grammar $G=(V,\Sigma,P,S)$, there exists an
  equivalent context free grammar $G'=(V',\Sigma,P',S)$
  such that the length of the right hand side of any 
  production belonging
  to $P'$ is at most $2$, $\var(G')\leq\symb(G)$, and the 
  rank of $G'$ coincides with the rank of $G$.
\end{lem}

\begin{pf}
  Without loss of generality, we suppose that for each variable
  $A$ in the set $V$ there is a production with $A$ on the
  left hand side.
  
  The set $V'$ of variables of $G'$ is defined by considering all
  variables in the set $V$, plus some extra variables as defined
  below. The set of productions $P'$ is defined as follows:
  We consider each production $A\rightarrow X_1X_2\ldots X_m$
  belonging to $P$, with $X_i\in V\cup\Sigma$, $i=1,\ldots,m$:
  \begin{itemize}
  \item If $m\leq 2$, then the production 
  $A\rightarrow X_1X_2\ldots X_m$ belongs to $P'$.
  \item If $m>2$, then $m-2$ new extra variables
  $D_1,D_2,\ldots,D_{m-2}$ are introduced in the set $V'$, and the
  following productions are added to $P'$:
  \[A\rightarrow X_1D_1, D_1\rightarrow X_2D_2, \ldots,
    D_{m-3}\rightarrow X_{m-2}D_{m-2},
    D_{m-2}\rightarrow X_{m-1}X_m.
  \]
  \item No other productions are in $P'$.
  \end{itemize}
  Note that the construction of $P'$ is similar to the last
  step in the classical reduction of general context-free grammars
  to Chomsky normal form.
  
  It is easy to verify that $G$ and $G'$ generate the same
  language. Furthermore, by construction, the right hand side of
  each production of $G'$ has length at most $2$.
  
  We recall that each variable of $V$ appears on the left hand
  side of some production. Furthermore, for each
  production $A\rightarrow\alpha$, with $|\alpha|>2$,
  $|\alpha|-2$ extra variables have been introduced in $V'$.
  Hence:
  \begin{eqnarray*}
  \symb(G)&  = &\sum_{(A\rightarrow\alpha)\in P}(2+|\alpha|)\\
          &\geq& \#V+\sum_{(A\rightarrow\alpha)\in P}(1+|\alpha|)\\
          &\geq& \#V+\sum_{(A\rightarrow\alpha)\in P {\rm\ s.t.\ } |\alpha|> 2}(1+|\alpha|)\\
          &\geq&\#V+\sum_{(A\rightarrow\alpha)\in P {\rm\ s.t.\ } |\alpha|> 2}(|\alpha|-2)\\
          & = & \var(G').
\end{eqnarray*}
Finally, it is immediate to observe that the definition of
$G'$ preserves the rank.
\hfill\qed
\end{pf}

\begin{cor}\label{cor:cfg2}
  Given an alphabet $\Sigma=\set{a_1,\ldots,a_m}$, for any context--free
  grammar $G$ with $\symb(G)=h$ and generating 
  a letter-bounded language $L\subseteq a_1^*\ldots a_m^*$, there exists an equivalent
  $(m-1)$-turn PDA $M$ with $2^{O(h)}$ states and $O(1)$ stack symbols.
\end{cor}

\begin{pf} 
At first, it can be observed that Lemma~\ref{lemma:der} is true not only for CFGs in Chomsky normal form but also for CFGs whose productions have right hand sides of length at most 2. Thus, all arguments in Section 3 are also true for such ``normalized'' CFGs. Owing to Lemma~\ref{lemma:norm}, we then observe that any CFG $G$ can be converted to an equivalent CFG $G'$ such that the length of the right hand side of any production belonging to $G'$ is at most 2 and $\var(G') \le \symb(G)$. With similar arguments as in Corollary~\ref{cor:cfg} we obtain the claim.
\qed
\end{pf}

We will discuss and present the extension of Corollary~\ref{cor:cfg2}
to the word-bounded case in Section~\ref{sec:wb}.

\section{Reducing the Number of Turns}
\label{sec:turns}

By the results presented in Section~\ref{sec:cfg}, each context-free
subset of $a_1^*\ldots a_m^*$ can be accepted by an $(m-1)$-turn PDA.
In particular, Corollary~\ref{cor:pda} shows that the size of an 
$(m-1)$-turn PDA equivalent to a given PDA of size $n$ accepting a subset of $a_1^*\ldots a_m^*$, is
at most exponential in the square of $n$.

In this section, we further deepen this kind of investigation by studying
how to convert an arbitrary $k$-turn PDA accepting a letter-bounded language $L \subseteq a_1^*a_2^*\ldots a_m^*$ to an equivalent $(m-1)$-turn PDA.
It turns out that the increase in size is at most polynomial.
All PDAs we consider are in normal form.

\medskip

Let us start by considering the unary case, i.e., $m=1$, which turns
out to be crucial to get the simulation in the general case.
In the following we consider a $k$-turn PDA $M=(Q,\Sigma,\Gamma,
\delta,q_0,Z_0,F)$ in normal form.
Then we know that at most one symbol is pushed on the stack in every transition.

\begin{lem} \label{lem:nfa}
Let $M$ be a PDA accepting a unary language $L$. Let $L(q_1,A,q_2)$ be the set of all words
which are processed by 1-turn sequences $\pi$ of
configurations starting with some stack height $h$ in a state $q_1$ and having $A$ as topmost stack
symbol
and ending with the same stack height $h$ in some state $q_2$ and having $A$ as topmost stack symbol.
Then,
$L(q_1,A,q_2)$ can be recognized by an NFA $M'$ such that $size(M') \le n^2$ and $n=size(M)$.
\end{lem}

\begin{pf}
Consider the following CFG $G$ with start symbol $[q_1,A,q_2]$ having the following
productions.
Let $p,p',q,q' \in Q$, $Z \in \Gamma$, and $\sigma \in \{a,\epsilon\}$.
\begin{enumerate}
\item $[p,Z,q] \rightarrow \sigma [p',Z,q]$, if $\delta(p,Z,\sigma) \ni (p',-)$,
\item $[p,Z,q] \rightarrow \sigma [p,Z,q']$, if $\delta(q',Z,\sigma) \ni (q,-)$,
\item $[p,Z,q] \rightarrow [p',Z',q']$, if $\delta(p,Z,\epsilon) \ni (p',push(Z'))$ and
$\delta(q',Z',\epsilon) \ni (q,pop)$,
\item $[p,Z,q] \rightarrow \epsilon$, if $p=q$.
\end{enumerate}

We want to describe how $M'$ simulates a 1-turn sequence $\pi$. 
We simulate the parts of $\pi$ with $A$ as topmost stack symbol and stack height $h$ with productions (1)
and (2). The first part from the beginning up to the first push operation is simulated using productions (1).
The second part starting at the end of the computation and going backwards up to the last pop operation
is simulated with productions (2).
We may change nondeterministically between productions (1) and (2). This is possible, since the input is
unary. Having simulated the parts of $\pi$ with stack height $h$ it is decided nondeterministically to
proceed with simulating the parts of $\pi$ with stack height $h+1$. Productions (3) simulate a push operation
and the corresponding pop operation. Then, productions (1) and (2) can be again used to simulate the
parts of $\pi$ with stack height $h+1$. Now, we iterate this behavior and simulate all computational
steps in $\pi$ while the stack height simulated is growing. Finally, we can terminate the derivation with productions (4) when the stack height has reached its highest level and all computational steps have been simulated.

Now, we construct an equivalent NFA $M'=(Q',\Sigma,\delta',(q_1,A,q_2),F')$ as follows:
$Q'=Q \times \Gamma \times Q$ and $F'=\{(q,Z,q) \mid q \in Q, Z \in \Gamma\}$.
For $\sigma \in \{a,\epsilon\}$ the transition function $\delta'$ is defined as:
\begin{enumerate}
\item $\delta'((p,Z,q),\sigma) \ni (p',Z,q)$, if $\delta(p,Z,\sigma) \ni (p',-)$,
\item $\delta'((p,Z,q),\sigma) \ni (p,Z,q')$, if $\delta(q',Z,\sigma) \ni (q,-)$,
\item $\delta'((p,Z,q),\sigma) \ni (p',Z',q')$, if $\delta(p,Z,\epsilon) \ni (p',push(Z'))$ and
$\delta(q',Z',\epsilon) \ni (q,pop)$,
\end{enumerate}
It is not difficult to observe that $T(M')=L(G)$.\qed
\end{pf}

\begin{cor}\label{cor:1turn}
Let $M$ be some 1-turn PDA accepting a unary language $L$. Then, an equivalent NFA $M'$ can be
constructed such that $size(M') \le n^2+1$ and $n=size(M)$.
\end{cor}

\begin{pf}
We can use the above construction, but additionally have to guess in a first step in which state a
computation ends. Therefore, we add a new start symbol $S$ and add productions $S \rightarrow [q_0,Z_0,q_f]$
for all $q_f \in F$. For the NFA construction we add a new initial state $q_0'$ and the following rules
$\delta'(q_0',\epsilon) \ni (q_0,Z_0,q_f)$, for all $q_f \in F$ to $M'$.

It is easy to observe that the parts of $\pi$ with stack height one can be again simulated with productions (1)
and (2). The remaining part of the simulation is identical to the above described construction.
\qed
\end{pf} 

A subcomputation $\pi'$ is called {\it strong} of level
$A$ if it starts with some stack height $h$ and topmost stack symbol $A$, ends with the same stack height $h$ and topmost stack symbol $A$, and in all other configurations of $\pi'$ the stack height is greater than $h$. 

\begin{lem} \label{lem:nfa:turns}
Let $M$ be some $k$-turn PDA accepting a unary language $L$. 
Let $L(q_1,A,q_2)$ be the set of all words which are processed by sequences $\pi$ of
strong computations of level $A$ which, additionally, start in some state $q_1$ and end in some state
$q_2$. It can be observed that all words in $L(q_1,A,q_2)$ are accepted with $j \le k$ turns.
Then, $L(q_1,A,q_2)$ can be accepted by an NFA $M'$ such that $size(M') \in O(n^{2\lfloor \log_2 j
\rfloor+2})$ and $n=size(M)$.
\end{lem}

\begin{figure} 
\centerline{\psfig{figure=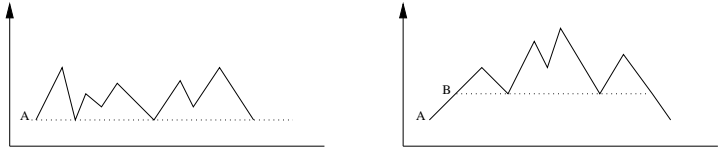,width=9.5cm}}
\caption{The two cases arising in the construction in Lemma~\ref{lem:nfa:turns}.} \label{fig:types}
\end{figure}

\begin{pf} 
The construction is very similar to the above described construction. Additionally, we store the number
of turns, which have to be simulated, in the fourth component of the variables. There are two cases
how $\pi$ may look like. In the first case (type I, cf. Fig.~\ref{fig:types}, left) $\pi$ consists of
at least two strong computations of level $A$. We introduce a new production type (5) which is used  
to decompose a sequence of strong computations with $i$ turns into two subsequences with $i_1$ and $i_2$
turns, respectively. A resulting subsequence is then either again of type I and can be again decomposed
with the new productions (5), or it is of type II, i.e., it consists of one strong computation of level $A$ (cf.
Fig.~\ref{fig:types}, right). If this computation is 1-turn, it can be simulated with the productions (1) to
(3) and finished with productions (4). If it is not 1-turn, we can reduce it to a sequence of strong
computations of level $B$ by using the productions (1) to (3). Then, the same analysis can be made for strong
computations of level $B$. 

The formal construction of the CFG $G$ is as follows. We consider the start symbol $[q_1,A,q_2,j]$ and the
following productions.
Let $p,p',q,q' \in Q$, $Z \in \Gamma$, and $\sigma \in \{a,\epsilon\}$.
\begin{enumerate}
\item $[p,Z,q,i] \rightarrow \sigma [p',Z,q,i]$, if $\delta(p,Z,\sigma) \ni (p',-)$,
\item $[p,Z,q,i] \rightarrow \sigma [p,Z,q',i]$, if $\delta(q',Z,\sigma) \ni (q,-)$,
\item $[p,Z,q,i] \rightarrow [p',Z',q',i]$, if $\delta(p,Z,\epsilon) \ni (p',push(Z'))$ and
$\delta(q',Z',\epsilon) \ni (q,pop)$,
\item $[p,Z,q,1] \rightarrow \epsilon$, if $p=q$,
\item $[p,Z,q,i] \rightarrow [p,Z,r,i_1][r,Z,q,i_2]$, for all $r \in Q$ and $i_1,i_2 \ge 1$ such that
$i_1+i_2 \le i$.
\end{enumerate}

It can be shown by an induction on the number of turns that $G$ generates $L(q_1,A,q_2)$.

We can observe that all productions are right-linear except for productions (5). Since the last component of a
variable $[p,Z,r,i]$ is reduced in every application of a production (5), we can conclude that (5) is applied
at most $j-1$ times.
Thus, every sentential form contains at most $j$ variables. Thus, we can construct some NFA
simulating the single derivation steps by representing all variables of a sentential form in its state. A rough estimation of the number of states is then
$$\sum_{i=1}^j (j|Q|^2|\Gamma|)^i = \frac{(j|Q|^2|\Gamma|)^{j+1} -1}{j|Q|^2|\Gamma|-1}-1 <
\frac{(j|Q|^2|\Gamma|)^{j+1}}{j|Q|^2|\Gamma|-1} \le 2(j|Q|^2|\Gamma|)^j \in O(n^{2j}).
$$ 

We now want to do some finer estimation and will obtain $2(j|Q|^2|\Gamma|)^{\lfloor \log_2 j \rfloor+1}$ as
upper bound. To this end, we observe that a simulation of a production (5) increases the number of variables
in the current state of the NFA by one and that a simulation of a production (4) at the end of some 1-turn
computation decreases the number of variables by one. Thus, our strategy is to apply productions of type
(4) as soon as possible. Now, whenever an application of a production (5) has replaced a variable
$[p,Z,q,i]$ by two variables $[p,Z,r,i_1]$ and $[r,Z,q,i_2]$, then the derivation of the
variable with the lower number of remaining turns is simulated. This makes sure that the total
number of variables in a state is as small as possible. The worst case which can occur in this
context is that in every application of a production (5) the number of turns is divided into two equal parts.
This may happen at most $\lfloor \log_2 j \rfloor+1$ many times. Thus, the size of the NFA can be
estimated as follows

\begin{multline*}
\sum_{i=1}^{\lfloor \log_2 j \rfloor+1} (j|Q|^2|\Gamma|)^i = \frac{(j|Q|^2|\Gamma|)^{\lfloor \log_2 j
\rfloor+2} -1}{j|Q|^2|\Gamma|-1}-1 < \frac{(j|Q|^2|\Gamma|)^{\lfloor \log_2 j
\rfloor+2}}{j|Q|^2|\Gamma|-1}=\\
= \frac{j|Q|^2|\Gamma|}{j|Q|^2|\Gamma|-1}(j|Q|^2|\Gamma|)^{\lfloor \log_2 j
\rfloor+1} \le 2(j|Q|^2|\Gamma|)^{\lfloor \log_2 j \rfloor+1} \in O(n^{2\lfloor
\log_2 j \rfloor+2})
\end{multline*}

Finally, it can be observed that the last component in a tuple $[p,Z,q,i]$ may be removed, since the
maximum number of possible turns only depends on $p,Z,q$. This may save the constant factor $j$ in the
above estimation.
\qed
\end{pf} 

\begin{cor} \label{cor:nfa}
Let $M$ be some $k$-turn PDA accepting a unary language $L$. Then, an equivalent NFA $M'$ can be
constructed with $size(M') \in O(n^{2\lfloor \log_2 k \rfloor+2})$ and $n=size(M)$.
\end{cor}

\begin{pf}
Observe that an accepting computation in $M$ is a sequence of strong computations of level $Z_0$ starting
in $q_0$ and ending in some accepting state.
\qed
\end{pf} 

Now, we are able to consider the general case, i.e., $m\geq 1$ and start with some definitions.

Given the alphabet $\Sigma=\set{a_1,\ldots,a_m}$, we define the set $\Pi(m)$ as follows 
$$
\Pi(m)=\{a_i \mid 1 \le i \le m\} \cup \{a_ia_j \mid 1 \le i < j \le m\}
$$

It is easy to show that the cardinality of $\Pi(m)$ is $\frac{m^2+m}{2}$. Let $w \in a_1^*a_2^* \ldots
a_m^*$ be some string. Then $\pi_l(w)$ denotes the projection to the first symbol of $w$ and $\pi_r(w)$
denotes the projection to the last symbol of $w$. For example, let $w=a_2a_3a_4$. Then, $\pi_l(w)=a_2$
and $\pi_r(w)=a_4$.

\begin{thm} \label{thm:kturn}
Let $M$ be some $k$-turn PDA accepting a letter-bounded language $L \subseteq a_1^*a_2^* \ldots a_m^*$. Then, an
equivalent $(m-1)$-turn PDA $M'$ can be constructed such that $size(M') \in O(m^6 n^{4\lfloor \log_2 k
\rfloor+8})$ and $n=size(M)$.
\end{thm}  

\begin{pf}
It has been shown in the previous section that any $L \subseteq a_1^*a_2^* \ldots a_m^*$ can be accepted
by an $(m-1)$-turn PDA. If $L$ is accepted by a $k$-turn PDA such that $k>m-1$, then some turns are in a
way ``not necessary.'' We will show in this proof that this finite number of additional
turns takes place within unary parts of the input, i.e., while reading some input $a_i^*$ with $1 \le i
\le m$. Then, with the help of the construction of Lemma~\ref{lem:nfa:turns}, these parts can be accepted by
NFAs and hence do not affect the stack height in the construction of an $(m-1)$-turn PDA accepting $L$.

The construction is similar to the constructions of the previous two lemmas. Additionally, we introduce a fifth
component of the variables in which some element $s \in \Pi(m)$ is stored. If $s=a_i$ ($1 \le i \le m$), this means that 
the variable can only produce terminals $a_i$. If $s=a_ia_j$ ($1 \le i<j \le m$), then such a variable can only produce sentential forms which start with terminals $a_i$ and end with terminals $a_j$.	

For the construction we first consider a context-free grammar $G$ with start symbol $S$ and having the following productions.
Let $p,p',q,q' \in Q$, $Z \in \Gamma$, $a,b \in \{a_1,\ldots,a_m\}$ such that $ab \in \Pi(m)$,
$\overline{a} \in \{a,\epsilon\}$, and $\overline{b} \in
\{b,\epsilon\}$.
\begin{enumerate}
\item $[p,Z,q,i,ab] \rightarrow \overline{a} [p',Z,q,i,ab]$, if $\delta(p,Z,\overline{a}) \ni (p',-)$,
\item $[p,Z,q,i,ab] \rightarrow [p,Z,q',i,ab] \overline{b}$, if $\delta(q',Z,\overline{b}) \ni (q,-)$,
\item $[p,Z,q,i,ab] \rightarrow [p',Z',q',i,ab]$, if $\delta(p,Z,\epsilon) \ni
(p',push(Z'))$ as well as $\delta(q',Z',\epsilon) \ni (q,pop)$, 
\item $[p,Z,q,1,u] \rightarrow \epsilon$, if $p=q$, for all $u \in \Pi(m)$,
\item $[p,Z,q,i,ab] \rightarrow [p,Z,r,i_1,u][r,Z,q,i_2,v]$, for all $r \in Q$, $i_1,i_2 \ge 1$ such that
$i_1+i_2 \le i$, $|u|>1$, $|v|>1$, $\pi_l(u)=a$, and $\pi_r(v)=b$,
\item $[p,Z,q,i,ab] \rightarrow ([p,Z,r,i_1,a],[r,Z,q,i_2,v])$, for all $r \in Q$, $i_1,i_2 \ge 1$ such
that $i_1+i_2 \le i$, $\pi_r(v)=b$, 
\item $[p,Z,q,i,ab] \rightarrow ([p,Z,r,i_1,u],[r,Z,q,i_2,b])$, for all $r \in Q$, $i_1,i_2 \ge 1$ such
that $i_1+i_2 \le i$, $\pi_l(u)=a$.
\item $S \rightarrow [q_0,Z_0,q_f,i,u]$ for all $q_f \in F$, $1 \le i \le k$, and $u \in \Pi(m)$,
\end{enumerate}

The productions (1) to (4), and (8) are defined similarly to the previous constructions. In the productions (5) to (7),
a computation with $i$ turns is decomposed into two subcomputations with $i_1$ and $i_2$ turns, respectively. Additionally,
we differentiate whether we obtain subcomputations producing only one type of terminals or not. The former case is handled with productions (6) and (7), for the latter case we have the productions (5).

Moreover, we have to define productions for variables of the form $[p,Z,q,i,a]$ with $a \in \{a_1,\ldots,a_m\}$.
Such variables are from now on called ``unary'' variables. 
Owing to Lemma~\ref{lem:nfa:turns} we know that the language $L(p,Z,q)$ can be accepted by some NFA $A$
having at most $2(i|Q|^2|\Gamma|)^{\lfloor \log_2 i \rfloor+1}$ states. This NFA can be converted to
some right-linear grammar $G_A$ with at most $2(i|Q|^2|\Gamma|)^{\lfloor \log_2 i \rfloor+1}$
variables. Now, the productions for a unary variable $[p,Z,q,i,a]$ with $a \in \{a_1,\ldots,a_m\}$ are
defined to be the productions of the corresponding right-linear grammar $G_A$.

In order to finally get a PDA making at most $m-1$ turns, we have introduced in the productions (6) and (7) some special
variables of the form $([p,Z,q,i,a],[p',Z',q',i',v])$ which, at a first glance, are not natural and
intuitive.

To derive such a variable $([p,Z,q,i,a],[p',Z',q',i',v])$, we first derive its first component
$[p,Z,q,i,a]$ with the above defined unary productions. Observe that the resulting productions are right-linear. 
For variables having the form $(\epsilon,[p',Z',q',i',v])$ we add productions
$(\epsilon,[p',Z',q',i',v])\rightarrow [p',Z',q',i',v]$. The
remaining second component $[p',Z',q',i',v]$ is then derived with the productions (1) to (7) if $|v|>1$ and
with the above defined productions otherwise.  

To derive a variable $([p,Z,q,i,u],[p',Z',q',i',b])$, we first derive its second component
$[p',Z',q',i',b]$ with the above defined unary productions. Observe that the resulting productions are left-linear.
We add productions of the form $([p,Z,q,i,u],\epsilon)\rightarrow [p,Z,q,i,u]$ and the remaining first
component $[p,Z,q,i,u]$ is then derived with the productions (1) to (7) if $|u|>1$ and with the above defined
productions otherwise. We would like to remark that variables of the form $([p,Z,q,i,a],[p',Z',q',i',b])$ are treated as in
the first case, i.e., we start to derive the first component with unary productions and then derive the second
component.

It can be observed that $G$ generates $T(M)$. 
Since all productions in $G$ are linear except those of type (5),
the number of variables occurring in a sentential form can only be increased by applications of productions of type (5).
It can be observed that the maximum number of variables introduced by productions of type (5) is bounded by the maximum number of decompositions of the string $a_1a_2 \ldots a_m$ into substrings $w_1,w_2, \ldots, w_l$ such that, for $1 \le t \le l$, $w_t=a_ia_j$ with $1 \le i < j \le m$ and $w_1w_2 \ldots w_l \in a_1^*a_2^* \ldots a_m^*$. It is easy to show that $l \le m-1$. Thus, every sentential form
contains at most $m-1$ variables which implies that $G$ is ultralinear of rank $m-1$. 

We next convert $G$ to some equivalent one-state PDA $M'$ using the standard conversion algorithm as
given for example in \cite{hopcroftullman}. Since the rank of $G$ is $m-1$,
it can be observed that the maximum number of variables on the stack is bounded by $m-1$. 
Furthermore, any decreasing of the stack starts by deleting some variable from the stack. This action
corresponds to an application of some production of type (4) in the grammar. Thus, the number of turns in $M'$
is bounded by the number of possible applications of productions of type (4) which is in turn bounded by the
number of variables on the stack. Thus, the number of turns in $M'$ is bounded by $m-1$. 

We now want to estimate the size of $M'$ which is bounded by the number of variables of $G$ and the
size of the alphabet $\Sigma$. The number of variables of type $[p,Z,q,i,u]$ is bounded by
$O(km^2|Q|^2|\Gamma|)$ and the number of variables resulting from the unary productions is bounded by $m
(2k|Q|^2|\Gamma|)^{\lfloor \log_2 k \rfloor+1}$. We now want to estimate the number of variables
resulting from variables of type $([p,Z,r,i,u],[r,Z,q,j,v])$. Observe that in each such
variable $|u|=1$ or $|v|=1$, respectively. Since these unary parts are derived first, the number of
variables resulting is bounded by the product of the number of variables resulting from unary
productions and of the number of variables of type $[p,Z,q,i,u]$. Thus, the number of variables
resulting from variables of type $([p,Z,r,i,u],[r,Z,q,j,v])$
is bounded by $O(m^3(2k|Q|^2|\Gamma|)^{\lfloor \log_2 k \rfloor+2})$ which implies $|M'| \in
O(m^3(2k|Q|^2|\Gamma|)^{\lfloor \log_2 k \rfloor+2})$.

Finally, we convert $M'$ to a PDA in normal form. This may cause at most an additional quadratic blow-up.
Thus, we obtain
$O(m^6(2k|Q|^2|\Gamma|)^{2\lfloor \log_2 k \rfloor+4})=O(m^6 n^{4\lfloor
\log_2 k \rfloor+8})$ as an upper bound.
\qed
\end{pf}

\begin{cor} \label{cor:kturn1}
The trade-offs between finite-turn pushdown automata that accept letter-bounded languages are at most
polynomial.
\end{cor}

\section{Lower Bounds}
\label{sec:lower}

In this section we show the optimality of the simulation of grammars generating letter-bounded languages by
finite-turn PDAs (Corollary~\ref{cor:cfg}), and of some other simulation results presented in the paper.
Even in this case, the preliminary investigation of the unary case will be useful to afford the general
case. 

\begin{thm} \label{lowerbound_unary}
For any integer $n \ge 1$, consider the language  $L_{n}=\{a^{2^n}\}$.
\begin{enumerate}
  \item $L_n$ can be generated by some CFG in Chomsky normal form with $n+1$ variables.
  \item Every NFA accepting $L_n$ needs at least $2^n$ states.
  \item For each $k>0$, every $k$-turn PDA accepting $L_{n}$ is at least
   of size $2^{cn}$ for some constant $c>0$ and any 
   sufficiently large $n$.
\end{enumerate}
\end{thm}

\begin{pf} 
To prove (1) it is enough to observe that $L_n$ can be generated by the 
grammar $G$ with the following productions:
\begin{eqnarray*}
S & \rightarrow & A_1A_1\\
A_1 & \rightarrow & A_2A_2\\
  & \vdots & \\
A_{n-1} & \rightarrow & A_nA_n\\
A_n & \rightarrow & a
\end{eqnarray*}

The proof of (2) is trivial.

Finally, to prove (3) consider a $k$-turn PDA $M$ of size  $s(n)$ accepting $L_n$. Due to
Corollary~\ref{cor:nfa} we can
construct an equivalent NFA of size $s'(n)\leq  Hs^K(n)$ for suitable constants $H,K$. Since $s'(n)\geq
2^n$, we obtain $s(n)\geq H'2^{n/K}$ for some other constant $H'$.
\qed
\end{pf}

{F}rom Theorem~\ref{lowerbound_unary}(3), it turns out that for each integer $m$ the simulation result
stated in Corollary~\ref{cor:cfg} is optimal.
The witness languages are unary. Hence, they can be also accepted 
by ``simpler'' devices, i.e., finite automata or PDAs with less than $m-1$ turns.
We now show the optimality in a stronger form, by exhibiting, for each integer $m$, a family of witness
languages that cannot be accepted with less than $m-1$ turns. 

\begin{thm} \label{lowerbound_m}
Given the alphabet $\Sigma=\set{a_1,\ldots,a_m}$, for
any integer $n \ge 1$ consider the language
\[
\tilde L_n=\{a_1^{n_0+n_1}a_2^{n_1+n_2}\ldots a_{m-1}^{n_{m-2}+n_{m-1}}a_m^{n_{m-1}}\mid n_0=2^n, n_1\geq
1,\ldots, n_{m-1}\geq 1\}.
\]
\begin{enumerate}
  \item $\tilde L_n$ is generated by some CFG in Chomsky normal form with $n+4m-3$ variables.
  \item $\tilde L_n$ is accepted by an $(m-1)$-turn PDA of size $2^{O(n)}$.
  \item For each integer $k\geq m-1$, every $k$-turn PDA accepting $\tilde L_{n}$ is at least of size
$2^{cn}$ for some constant $c>0$ and any sufficiently large $n$.
  \item $\tilde L_{n}$ cannot be accepted by any PDA which makes less than $m-1$ turns.
\end{enumerate}
\end{thm}

\begin{pf}
Consider the grammar $G$ with the following productions:
\begin{itemize}
\item $S \rightarrow  A_0B_1$,
\item $B_1 \rightarrow C_1B_2,\ldots, B_{m-3} \rightarrow  C_{m-3}B_{m-2},
B_{m-2} \rightarrow  C_{m-2}C_{m-1}$
\item $A_0 \rightarrow A_1A_1, A_1  \rightarrow A_2A_2, \ldots,
A_{n-1} \rightarrow A_nA_n, A_n  \rightarrow a_1$
\item $C_i \rightarrow D_iE_i \mid D_iD_{i+1}, E_i \rightarrow C_iD_{i+1}$, 
for $i=1,\ldots,m-1$ 
\item $D_i \rightarrow a_i$, for $i=1,\ldots,m$.
\end{itemize}
\noindent
It is possible to verify that this grammar $G$ generates the language 
$\tilde L_n$. In particular, observe that from each $C_i$ we can derive 
terminal strings only of the form $a_i^{t}a_{i+1}^t$, with $t\geq 1$,
and from $A_0$ we can derive only the string $a_1^{2^n}$. 
By observing that we can use the same 
variable for $A_n$ and $D_1$, we easily conclude that the total number of variables is $n+4m-3$.
This proves (1). Furthermore, as an easy consequence, applying 
Corollary~\ref{cor:cfg} we get (2).

Now, given $k\geq m-1$, suppose to have a $k$-turn PDA of size $s$ accepting $\tilde L_n$. By replacing
each move consuming
a symbol $a_i$, where $i>1$, with an $\epsilon$-move, we get another 
$k$-turn PDA with $s$ states accepting the
language $\tilde L'_n=\{a_1^{n_1}\mid n_1>2^n\}$. 
Using a slight modification of Theorem~\ref{lowerbound_unary}, we
can get that $s\geq 2^{cn}$ for some constant number $c>0$ and 
any sufficiently large $n$. This proves (3).

We finally prove (4). To this aim, we first prove that each 
context-free grammar which generates the
following language $L$ must have rank at least $m-1$:
\[
L=\{a_1^{n_1}a_2^{n_1+n_2}\ldots a_{m-1}^{n_{m-2}+n_{m-1}}a_m^{n_{m-1}}\mid 
n_1\geq 1,\ldots, n_{m-1}\geq 1\}.
\]
Let $G$ be a grammar with $h$ variables which generates $L$. 
We can suppose that the right hand side of each production of
$G$ has length at most $2$ (by Lemma~\ref{lemma:norm} this
restriction preserves the rank, furthermore it can be easily
seen that Lemma~\ref{lemma:der}, used in the following,
holds even for there grammars).
Let $H=2^h$ and $z=a_1^Ha_2^{2H}\ldots a_{m-1}^{2H}a_m^H\in L$.

Given a derivation tree $T:S\genera z$, consider an integer $k>0$,
derivation trees $T_0,\ldots,T_k$ of strings $z_0,\ldots,z_k$, partial 
derivation trees $U_i:A_i\Genera v_iA_ix_i$, $i=1,\ldots,k$, according
to Lemma~\ref{lemma:der}.
For $i=1,\ldots,k$, let $border(A_i)=(l_i,r_i)$. 
Note that $r_i=l_i+1$, otherwise, by pumping $T_{i-1}$ (which generates the string $z_{i-1}\in L$)
with the partial tree $U_i$, the resulting string $z_i$ 
should not belong to $L$.
Considering the definition of the relation $\leq$ between borders and
Lemma~\ref{lemma:border}, this easily implies that for each
pair of variables $A_i,A_j\in\set{A_1,\ldots A_k}$, either
$A_i$ and $A_j$ have the same border or they lie on two different paths from the root of the tree $T$.
We now prove that for each $j=1,\ldots,m-1$, there is a variable
$A_{i_j}\in\set{A_1,\ldots,A_k}$ such that $border(A_{i_j})=(j,j+1)$,
obtaining in this way $m-1$ variables
belonging to different paths from the root of $T$.

Suppose, by contradiction, that there is an index $\tilde\jmath$
such that $(\tilde\jmath,\tilde\jmath+1)\notin\set{border(A_i)\mid i=1,\ldots,k}$. 
Hence, there is an index $r$, $1\leq r\leq k$, such that
$0<l_1\leq\ldots\leq l_r\leq\tilde\jmath-1$ and $l_{r+1}>\tilde\jmath$
($r=1$ in the case $\tilde\jmath=1$).
The pumping process described in Lemma~\ref{lemma:der} starts from
the tree $T_0$, which generates a string 
$z_0=a_1^{n_1}a_2^{n_1+n_2}\ldots a_{m-1}^{n_{m-2}+n_{m-1}}a_m^{n_{m-1}}$. The number of occurrences of
the letters 
$a_1,\ldots,a_{\tilde\jmath}$ can be incremented only by pumping with the partial trees $U_1,\ldots,U_r$,
while the number of occurrences of
the letters $a_{\tilde\jmath+1},\ldots,a_m$ only by pumping with the
partial trees $U_{r+1},\ldots,U_k$.
Hence, the terminal string generated at the $r$th step should be
$$z_r=a_1^Ha_2^{2H}\ldots a_{\tilde\jmath-1}^{2H}a_{\tilde\jmath}^{2H}
a_{\tilde\jmath+1}^{n_{\tilde\jmath}+n_{\tilde\jmath+1}}\ldots a_{m-1}^{n_{m-2}+n_{m-1}}a_m^{n_{m-1}}.$$
For $i=1,\ldots,\tilde\jmath$, let $\alpha_i$ be the number of
occurrences of letters $a_i$ and $a_{i+1}$ added during
the pumping process, which leads from $z_0$ to $z_r$,
by those partial trees among $U_1,\ldots,U_r$ such that the
borders of their roots coincide with $(i, i+1)$.
It is easy to verify that $\alpha_i=H-n_i$, for
$i=1,\ldots,\tilde\jmath$. Furthermore, by our choice of
$\tilde\jmath$, it turns out that $\alpha_{\tilde\jmath}=0$.
This implies that $n_{\tilde\jmath}=H$. This is a
contradiction, because $n_{\tilde\jmath}<|z_0|<H$.
Hence, we finally get that the tree $T$ contains $m-1$ variables 
$A_{i_1},\ldots,A_{i_{m-1}}\in\set{A_1,\ldots,A_k}$ which lie on
different paths from the root, i.e., there is a derivation of the form
$S\genera \alpha_1A_{i_1}\alpha_2\ldots\alpha_{m-1}A_{i_{m-1}}\alpha_m\genera z$.
Thus, we conclude that the rank of the grammar $G$ is at least $m-1$.

\smallskip
Using a slight modification of the construction given in the proof 
of Theorem~\ref{thm:kturn}, we can show that
from a $k$-turn PDA it is possible to get an equivalent grammar of 
rank $k$.
This implies that if a $k$-turn PDA accepts $L$ then $k\geq m-1$.

\smallskip
To complete the proof, we observe that given a PDA $\tilde M$ accepting
the language $\tilde L$, we can build a PDA $M$ which accepts $L$
by working in two phases: 
In the first phase $M$ simulates the moves of $\tilde M$ from the initial
configuration, as long as $\tilde M$ consumes the input $a_1^{2^n}$. 
In this phase,
each move consuming the symbol $a_1$ is replaced by an $\epsilon$-move
(an internal variable counts, up to $2^n$, the number of these moves).
In this way, $M$ is able to reach, without consuming any input symbol,
every configuration reachable by $\tilde M$ by consuming the input
prefix $a_1^{2^n}$. At this point, the second phase can start. In this
phase $M$ makes exactly the same moves as $\tilde M$. It is easy to
see that $M$ accepts the language $L$. Furthermore, if the given
PDA $\tilde M$ is $k$-turn, $M$ is $k$-turn, too.

In conclusion, having proved that $L$ cannot be accepted by
$k$-turn PDAs with $k<m-1$, we can conclude that $\tilde L$ cannot
 be accepted by
$k$-turn PDAs with $k<m-1$, too.
\qed
\end{pf}

Remark that we have considered so far only CFGs in Chomsky normal form and the measure $\var$. It is easy to observe that we also obtain exponential trade-offs when considering the measure $\symb$. This shows that the result of Corollary~\ref{cor:cfg2} is also optimal. Since $\tilde L_{n}$ can be accepted by a PDA of size $O(n)$, we obtain that the result of Corollary~\ref{cor:pda} is nearly optimal.

We complete this section by considering again the unary case.
In particular, we prove that the upper bound stated in Corollary~\ref{cor:1turn}
is tight.

\begin{thm} \label{unary:1turn}
Consider the language family 
$$L'_{n}=\{a^t \mid t \ge 0 \wedge t \equiv 0 \mod n \wedge t \equiv 0 \mod n+1\}$$ 
for natural numbers $n \ge 2$. Then each
$L'_{n}$ can be accepted by some 1-turn PDA of size $2n+1$, but every NFA accepting $L'_{n}$
needs at least $n^2+n$ states.
\end{thm}

\begin{pf}
A 1-turn PDA accepting $L'_{n}$ starts with checking whether the length of the input is divisable by
$n$ in its states. At the same time, the input is stored in the stack. Then the PDA guesses that the
whole input is read and checks whether the length of the input, which is stored on the stack, is
divisable by $n+1$. Finally, the PDA accepts if the whole input is read, divisable by $n$ and $n+1$, and
the stack is empty. Otherwise, the input is rejected. It can be observed that such a PDA is 1-turn, has
one stack symbol (apart from $Z_0$) and has $2n+1$ states.
Since $\gcd(n,n+1)=1$, we can apply a result from \cite{holzerkutrib} and obtain the latter claim.
\qed
\end{pf}

\section{Word-Bounded Languages}
\label{sec:wb}

In this section we study how to extend our results from
the letter-bounded case to the word-bounded case.
The idea is that of reducing the latter case to
the former one. To this aim, a large part of the section
is devoted to prove that for fixed $m$ words 
$w_1,w_2,\ldots,w_m\in\Sigma^*$, 
$m$ symbols $a_1,\ldots,a_m$ and the homomorphism
$\phi$ associating with each symbol $a_i$ the string
$w_i$, $i=1,\ldots,m$, and for each context-free grammar
$G=(V,\Sigma,P,S)$ in Chomsky normal form generating 
a subset of $w_1^*w_2^*\ldots w_m^*$, we can get another 
context-free
grammar $\hat G$ whose number of symbols is linear in
the number of symbols of $G$, namely $\symb(\hat G) = O(\symb(G))$, 
and such that $L(\hat G)=\phi^{-1}(L(G))$,
i.e., for all integers $k_1,\ldots,k_m\geq 0$: 
$a_1^{k_1}\ldots a_m^{k_m}\in L(\hat G)$ if and only
if $w_1^{k_1}\ldots w_m^{k_m}\in L(G)$.

\medskip

The construction is given in two steps: first we introduce a new
grammar $G'=(V',\Sigma,P',S')$ equivalent to $G$. In such a grammar, the variables of $G$ are marked with some indices which are
useful to recognize where, in a derivation, a variable can produce
the first symbol of one of the $w_i$'s. 
This will be useful to get from $G'$, in a second step, the required
grammar $\hat G$.

We start by considering the following set of variables:
\[
V' = \set{[A,i,l,r,j]\mid A\in V, 1\leq l\leq r\leq m, 
0\leq i\leq|w_l|, 0\leq j\leq|w_r|}.
\]

The definition is given in such a way that a variable $[A,i,l,r,j]$ can 
generate all terminal strings of the form $\alpha\beta\gamma$
generated by $A$, such that $\beta\in w_l^*\ldots w_r^*$, $\alpha$ is
the suffix of $w_l$ which starts in position $i+1$ and $\gamma$
is the prefix of $w_r$ which ends in position $j$. If $l=r$, furthermore,
the variable $[A,i,l,l,j]$ will be able to generate the factor of
$w_l$ from position $i+1$ to position $j$ if $A$ is able to do this.

To this aim, we define the following productions:
\begin{enumerate}
\item $[A,j-1,l,l,j]\rightarrow a$,\\
  for all $A\in V$ such that $A\rightarrow a$ is a production
  in $P$, $1\leq l\leq m$, $1\leq j\leq|w_l|$, and 
  $a=w_{l,j}$, i.e., $a$ is the symbol in position $j$ of $w_l$.
\item $[A,i,l,r,j]\rightarrow [B,i,l,h,k][C,k,h,r,j]$,\\
  for all $A,B,C\in V$ such that $A\rightarrow BC$ 
  is a production in $P$, $1\leq l\leq h\leq r\leq m$, 
  $0\leq i\leq|w_l|$, $0\leq k\leq|w_h|$, and $0\leq j\leq|w_r|$.
\item $[A,i,l,r,0]\rightarrow [A,i,l,r,|w_r|]$,\\
  for all $1\leq l\leq r\leq m$, $0\leq i\leq|w_l|$.
\item $[A,|w_l|,l,r,j]\rightarrow [A,0,l,r,j]$,\\
  for all $1\leq l\leq r\leq m$, $0\leq j\leq|w_r|$.
\item $[A,i,l,r,0]\rightarrow [A,i,l,h,0]$,\\
  for all $1\leq l\leq h< r\leq m$, $0\leq i\leq|w_l|$.
\item $[A,|w_l|,l,r,j]\rightarrow [A,i,|w_h|,h,r,j]$,\\
  for all $1\leq l< h\leq r\leq m$, $0\leq j\leq|w_r|$.
\end{enumerate}

The initial symbol $S'$ of $G'$ is $[S,|w_1|,1,m,0]$.

The following lemma states the main property of the variables of the
grammar $G'$ and it is crucial in order to prove that $G'$ is
equivalent to $G$:

\begin{lem}
For each variable $[A,i,l,r,j]$ of the grammar $G'$ above
defined and for each string $w\in\Sigma^*$, 
it holds that $[A,i,l,r,j]\generaP w$
if and only if $A\generaG w$, and
\begin{itemize}
\item either $w=w_{l,i+1}\ldots w_{l,|w_l|}\beta w_{r,1}\ldots w_{r,j}$, with
  $\beta\in w_l^*\ldots w_r^*$,
\item or $l=r$, $i<j$, and $w=w_{l,i+1}\ldots w_{l,j}$.
\end{itemize}
\end{lem}
\begin{pf}
We first prove the \emph{only if} part.
Let $s\geq 1$ be an integer such that $[A,i,l,r,j]\GENERAP{s} w$.
The proof is by induction on $s$.

If $s=1$ then the derivation consists only of the production
$[A,j-1,l,l,j]\rightarrow w_{l,j}$.
This implies that $i=j-1$, $l=r$, and $A\generaG w_{l,j}$.

If $s>1$ then we have to consider several possibilities depending
on which kind of production is used in the first step
of the derivation. (The following list refers to the
above defined productions (2), \ldots, (6):)

\begin{enumerate}
\refstepcounter{enumi}
\item $[A,i,l,r,j]\rightarrow [B,i,l,h,k][C,k,h,r,j]$:\\
  Hence $[B,i,l,h,k]\generaP w'$, $[C,k,h,r,j]\generaP w''$
  where $w=w'w''$. We observe that, by induction hypothesis, 
  $B\generaG w'$,
  $C\generaG w''$, and then $A\generaG w$. The proof can be
  easily completed by considering the following subcases:
  \begin{itemize}
  \item $w'=w_{l,i+1}\ldots w_{l,|w_l|}\beta'w_{h,1}\ldots w_{h,k}$,
    $w''=w_{h,k+1}\ldots w_{h,|w_h|}\beta''w_{r,1}\ldots w_{r,j}$,
    where $\beta'\in w_l^*\ldots w_h^*$ and 
    $\beta'\in w_h^*\ldots w_r^*$.
    By choosing $\beta=\beta'w_l\beta''$, it turns out that
    $\beta\in w_h^*\ldots w_r^*$  and 
    $w=w_{l,i+1}\ldots w_{l,|w_l|}\beta w_{r,1}\ldots w_{r,j}$.
  \item $l=h=r$, $i<k<j$, $w'=w_{l,i+1}\ldots w_{l,k}$, and 
    $w''=w_{l,k+1}\ldots w_{l,j}$. Then $w=w_{l,i+1}\ldots w_{l,j}$.
  \item $l=h$, $i<k$, $w'=w_{l,i+1}\ldots w_{l,k}$, and 
    $w''=w_{l,k+1}\ldots w_{l,|w_h|}\beta w_{r,1}\ldots w_{r,j}$, 
    or $h=r$, $k<j$,
    $w'=w_{l,i+1}\ldots w_{l,|w_l|}\beta w_{r,1}\ldots w_{r,k}$,
    and $w''=w_{r,k+1}\ldots w_{r,j}$. 
    In both these cases, we get that
    $w=w_{l,i+1}\ldots w_{l,|w_l|}\beta w_{r,1}\ldots w_{r,j}$.
  \end{itemize}

\item $[A,i,l,r,0]\rightarrow [A,i,l,r,|w_r|]$:\\
  Hence $[A,i,l,r,|w_r|]\GENERAP{s-1} w$. By induction
  hypothesis we get that $A\generaG w$. 
  If $w=w_{l,i+1}\ldots w_{l,|w_l|}\beta' w_{r,1}\ldots 
  w_{r,|w_r|}$, with $\beta'\in w_l^*\ldots w_r^*$, then
  $w=w_{l,i+1}\ldots w_{l,|w_l|}\beta$, with 
  $\beta=\beta'w_r\in w_l^*\ldots w_r^*$.
  Otherwise, $l=r$ and then $w=w_{l,i+1}\ldots w_{l,|w_l|}\beta$,
  with $\beta=\epsilon$.
  
\item $[A,|w_l|,l,r,j]\rightarrow [A,0,l,r,j]$:\\
  This case is similar to the previous one.

\item $[A,i,l,r,0]\rightarrow [A,i,l,h,0]$:\\
  Hence $[A,i,l,h,0]\GENERAP{s-1} w$ and, by induction
  hypothesis, $A\generaG w$ and 
  $w=w_{l,i+1}\ldots w_{l,|w_l|}\beta$ for some 
  $\beta\in w_l^*\ldots w_h^*$. From $h<r$, it turns
  out that $\beta\in w_l^*\ldots w_r^*$.

\item $[A,|w_l|,l,r,j]\rightarrow [A,|w_h|,h,r,j]$:\\
  Hence $[A,|w_h|,h,r,j]\GENERAP{s-1} w$ and, by induction
  hypothesis, $A\generaG w$ and 
  $w=\beta w_{r,1}\ldots w_{r,j}$ for some 
  $\beta\in w_h^*\ldots w_r^*$. From $l<r$, it turns
  out that $\beta\in w_l^*\ldots w_r^*$.
\end{enumerate}

\bigskip\noindent
We now prove the \emph{if} part.
Even in this case the proof is by induction on
the number $s$ of the steps of the derivation $A\GENERAG{s}w$
under consideration. 

If $s=1$ then $|w|=1$ and $A\rightarrow w$ must be a production
in $P$. We have to consider two cases:
\begin{itemize}
\item $w=w_{l,i+1}\ldots w_{l,|w_l|}\beta w_{r,1}\ldots w_{r,j}$, 
  with $\beta\in w_l^*\ldots w_r^*$:\\
  Since $|w|=1$, it turns out that:
  \begin{itemize}
  \item either $w=w_{l,|w_l|}$, $\beta=\epsilon$,
        $i=|w_l|-1$, and $j=0$, or 
  \item $w=w_{r,1}$, $\beta=\epsilon$,$i=|w_l|$, and $j=1$, or
  \item $\beta=w=w_h$, for some $l\leq h\leq r$,
        $i=|w_l|$, and $j=0$.
  \end{itemize}
  In the first case, the grammar $G'$ should contain the
  production $[A,|w_l|-1,l,l,|w_l|]\rightarrow w$. Hence, 
  using productions (5), (3), and (1), we get:
  \begin{eqnarray*}
  [A,i,l,r,j]=[A,|w_l|-1,l,r,0]&\RightP&[A,|w_l|-1,l,l,0]\\
  &\RightP&[A,|w_l|-1,l,l,|w_l|]\RightP{w}.
  \end{eqnarray*}
  The second case is similar.
  In the third case, the grammar $G'$ contains, by construction,
  the production $[A,0,h,h,1]\rightarrow w$. Hence, using
  productions (6), (5), (3), (4), we get the following
  derivation:
  \begin{eqnarray*}
  [A,i,l,r,j]=[A,|w_l|,l,r,0]&\RightP&[A,|w_h|,h,r,0]\\
  &\RightP&[A,|w_h|,h,h,0]\RightP[A,|w_h|,h,h,1]\\
  &\RightP&[A,0,h,h,1]\RightP w.
  \end{eqnarray*}

\item $l=r$, $i<j$, and $w=w_{l,i+1}\ldots w_{l,j}$:\\
  This case is trivial.

\end{itemize}

\medskip

Now suppose $s>1$. 
The first production applied in the derivation of
$w$ should be $A\rightarrow BC$, for two variables $B$ and $C$,
such that $B\generaG w'$, $C\generaG w''$, and $w=w'w''$.
As before, the proof is divided into two cases:
\begin{itemize}
\item $w=w_{l,i+1}\ldots w_{l,|w_l|}\beta w_{r,1}\ldots w_{r,j}$, 
  with $\beta\in w_l^*\ldots w_r^*$:\\
We have to consider the following three possibilities:

\begin{itemize}
\item $w'=w_{l,i+1}\ldots w_{l,|w_l|}\beta'w_{h,1}\ldots w_{h,k}$,
    $w''=w_{h,k+1}\ldots w_{h,|w_h|}\beta''w_{r,1}\ldots w_{r,j}$,
    for some $l\leq h\leq r$, $0\leq k\leq|w_h|$, 
    $\beta'\in w_l^*\ldots w_k^*$, and $\beta''\in w_k^*\ldots w_r^*$.
    By the inductive hypothesis, we obtain
    $[B,i,l,h,k]\generaP w'$, $[C,k,h,r,j]\generaP w''$.
    Hence, $[A,i,l,r,j]\generaP w$.
    
\item $w'=w_{l,i+1}\ldots w_{l,k}$, 
    $w''=w_{l,k+1}\ldots w_{l,|w_l|}\beta w_{r,1}\ldots w_{r,j}$,
    for some $i<k\leq |w_l|$, $\beta\in w_l^*\ldots w_r^*$.
    By the inductive hypothesis, we obtain
    $[B,i,l,l,k]\generaP w'$, $[C,k,l,r,j]\generaP w''$.
    Hence, $[A,i,l,r,j]\generaP w$.

\item $w'=w_{l,i+1}\ldots w_{l,|w_l|}\beta w_{r,1}\ldots w_{r,k}$ 
    $w''=w_{r,k+1}\ldots w_{r,j}$,
    for some $0\leq k\leq j$, $\beta\in w_l^*\ldots w_r^*$.
    By the inductive hypothesis, we obtain
    $[B,i,l,r,k]\generaP w'$, $[C,k,r,r,j]\generaP w''$.
    Hence, $[A,i,l,r,j]\generaP w$.
\end{itemize}

\item $l=r$, $i<j$, and $w=w_{l,i+1}\ldots w_{l,j}$:\\
There exists an index $k$, $i<k<j$, such
that $w'=w_{l,i+1}\ldots w_{l,k}$ and $w''=w=w_{l,k+1}\ldots w_{l,j}$.
By the induction hypothesis, we obtain
$[B,i,l,l,k]\generaP w'$ and $[C,k,l,l,j]\generaP w''$. Hence,
$[A,i,l,l,j]\generaP w$.
\qed
\end{itemize}
\end{pf}

\begin{thm}
The context-free grammar $G'$ is equivalent to $G$.
Furthermore, for each string $x\in L(G)$,
$x=x_1\ldots x_n$, with $x_k\in\Sigma$ for $k=1,\ldots,n$,
there exists a derivation
$S'\generaP[A_1,j_1-1,l_1,l_1,j_1]\ldots[A_n,j_n-1,l_n,l_n,j_n]$,
for suitable indices $l_1,\ldots,l_n$, $j_1,\ldots,j_n$,
such that $[A_k,j_k-1,l_k,l_k,j_k]\rightarrow x_k$, for $k=1,\ldots,n$.
\end{thm}
\begin{pf}
Given $x\in w_1^*\ldots w_m^*$, as a consequence of the previous lemma,
it is immediate to conclude that $S\generaG x$ if and only if
$S'\generaP x$. Each symbol $x_k$ of the string $x$ will be generated in a
derivation $\theta:S'\generaP x$ using a production of the form
$[A_k,j_k-1,l_k,l_k,j_k]\rightarrow x_k$, for suitable $A_k, j_k, l_k$.
Hence, from the derivation $\theta$ it is easy to get a derivation
$\theta':S'\generaP [A_1,j_1-1,l_1,l_1,j_1]
\ldots[A_n,j_n-1,l_n,l_n,j_n]$.\qed
\end{pf}

At this point we are able to define the grammar $\hat G$.
The set of variables is the same as of $G'$, the set of production
$\hat P$ contains all the productions in $P'$, with the exception
of each production of the form
$[A,j-1,l,l,j]\rightarrow w_{l,j}$,
which is replaced with the production 
$[A,j-1,l,l,j]\rightarrow a_l$ for $j=1$, and
with the production $[A,j-1,l,l,j]\rightarrow\epsilon$
in the other cases.
It is immediate to prove that the language generated by $\hat G$ coincides with $\phi^{-1}(L(G))$.

Hence, we get the following result:

\begin{cor}\label{cor:word}
Given $m$ strings $w_1,\ldots,w_m$, $m$ symbols $a_1,\ldots,a_m$,
and the homomorphism $\phi$ associating with each symbol $a_i$ the string
$w_i$, $i=1,\ldots,m$, for each context-free grammar $G$ 
generating a word bounded language $L\subseteq w_1^*\ldots w_m^*$, 
there exists another context-free grammar $\hat G$ such that 
$L(\hat G)=\phi^{-1}(L(G))$, and $\symb(\hat G)=O(\symb(G))$.
\end{cor}

We are now ready to extend Corollary~\ref{cor:cfg2} to word-bounded
languages:
\begin{thm} \label{thm:word}
  Given $m$ strings $w_1,\ldots,w_m$, for any context--free
  grammar $G$ with $\symb(G)=h$ and generating 
  a word-bounded language $L\subseteq w_1^*\ldots w_m^*$, there exists 
  an equivalent
  $(m-1)$-turn PDA $M$ with $2^{O(h)}$ states and $O(1)$ stack symbols.
\end{thm}
\begin{pf}
  By Corollary~\ref{cor:word}, given $m$ symbols $a_1,\ldots,a_m$,
  we can get a context-free grammar $\hat G$ with 
  $\symb(\hat G)=O(\symb(G))$,
  which generates the language $\phi^{-1}(L(G))$.
  Using Corollary~\ref{cor:cfg2}, we are able to find an
  $(m-1)$-turn PDA $\hat M$ with $2^{O(\symb(\hat G))}$ states 
  and $O(1)$ stack symbols, recognizing $\phi^{-1}(L(G))$.
  
  {F}rom $\hat M$ we now define a PDA $M$ accepting $L$: 
  $M$ should simulate
  $\hat M$ with the difference that a move of $\hat M$ consuming the input
  symbols $a_i$ is simulated by a sequence of moves of $M$ consuming
  the input factor $w_i$. It is easy to implement this with $2^{O(h)}$
  states and without increasing the number of stack symbols.
\qed
\end{pf}

Because the letter-bounded case is a special case of the word-bounded
case, by the results presented in Section~\ref{sec:lower},
the upper bound presented in Theorem~\ref{thm:word} is tight.

Also Corollary~\ref{cor:kturn1} can be extended to the word-bounded case.

\begin{thm} \label{thm:word-kturn}
The trade-offs between finite-turn pushdown automata that accept word-bounded languages are at most polynomial.
\end{thm}
\begin{pf}
Given $m$ strings $w_1,\ldots,w_m$, let $M$ be some $k$-turn PDA 
accepting a word-bounded language 
$L \subseteq w_1^*w_2^* \ldots w_m^*$. 
We consider $m$ letters $a_1,\ldots,a_m$ and the morphism
$\phi$ associating with each $a_i$ the string $w_i$, 
$i=1,\ldots,m$.
{F}rom the given PDA $M$, a $k$-turn PDA $\hat M$ accepting
$\phi^{-1}(L)$ can be defined by simulating $M$ step by step,
with the only difference that the input tape of $M$
is replaced by a buffer of length $\max\{|w_i|\mid i=1,\ldots,m\}$.
When $\hat M$ has to simulate a move of
$M$ that reads an input symbol, it uses the next symbol from the buffer.
However, if the buffer is empty, then $\hat M$ reads 
the next symbol $a_i$ from its input tape and puts the
corresponding string $w_i$ in the buffer.
Note that the size of $\hat M$ is polynomial in the size of $M$.
This is essentially the construction for inverse homomorphism as is
given, e.g., in \cite{hopcroftullman}.

Now, using Theorem~\ref{thm:kturn}, from $\hat M$ it is possible
to get an equivalent $(m-1)$-turn PDA $\hat M'$, which
is still polynomial in the size of $M$. Finally, using the
same construction outlined in the proof of Theorem~\ref{thm:word},
this last PDA can be converted in another $(m-1)$-turn PDA $M'$,
accepting the original language $L$, whose size 
is polynomial in the size of $M$.
\qed
\end{pf}

\section{Conclusion}

In this paper, we have considered context-free grammars generating letter-bounded as well as word-bounded languages. We have shown that such languages can be accepted by finite-turn pushdown automata. Furthermore,
we have given constructions for converting context-free grammars which generate bounded languages to equivalent finite-turn pushdown automata as well as minimizing the number of turns of a given pushdown automaton accepting a bounded language. The resulting trade-offs concerning the size of description of the corresponding context-free grammars and pushdown automata have been shown to be exponential when starting with an arbitrary context-free grammar or an arbitrary pushdown automaton. When starting with a finite-turn pushdown automaton,
a polynomial trade-off has been obtained. Both trade-offs are in strong contrast to the non-bounded case where non-recursive trade-offs are known to exist. Moreover, the existence of non-recursive trade-offs implies that such conversion algorithms and minimization algorithms cannot exist in general. Additionally, equivalence or inclusion problems are undecidable for arbitrary context-free languages.

We have shown that boundedness is a structural limitation on context-free languages which reduces non-recursive trade-offs to recursive trade-offs.
Together with the known positively decidable questions such as equivalence or inclusion of bounded context-free languages, we obtain that context-free grammars and pushdown automata for bounded languages are much more manageable from a practical point of view than in the general case.

\end{document}